%% file: paperDI.tex
\documentclass[pra,aps,reprint,groupedaddress]{revtex4-1}
\usepackage{amsfonts}
\usepackage{amsmath}
\usepackage{float}
\usepackage{graphicx}
\usepackage{epstopdf}

\usepackage{verbatim}% gives the \begin{comment} \end{comment} commands 
\usepackage{color}
\usepackage[english]{babel}

\usepackage[bookmarksopen=true,colorlinks,linkcolor = green,backref]{hyperref}
\newcommand{\LL }{\langle \! \langle}
\newcommand{\rr }{\rangle \! \rangle}
\newcommand{\mycdot}{ \times }

\newcommand{\tempsubsection}[1]{}

\newcommand{\commAS}[1]{{\color{red} \bf [AS: #1 ]}}
\newcommand{\commAZ}[1]{{\color{cyan} \bf [AZ: #1 ]}}

\input{my_commands}

\begin{document}

\title{Double photo-electron momentum spectra of Helium at infrared wavelength}

\author{Alejandro Zielinski}
%\affiliation{Physics Department, Ludwig Maximilians Universit\"at, D-80333 Munich, Germany}

\author{Vinay Pramod Majety}
%\affiliation{Physics Department, Ludwig Maximilians Universit\"at, D-80333 Munich, Germany}

\author{Armin Scrinzi}
\affiliation{Physics Department, Ludwig Maximilians Universit\"at, D-80333 Munich, Germany}

\date{\today}

%\tableofcontents{}

\begin{abstract}
Double photo-electron momentum spectra of the Helium atom  are calculated \textit{ab initio}
at extreme ultra-violet and near infrared wavelengths. At short wavelengths two-photon
double ionization yields, two-electron energy spectra, and triply differential cross sections 
agree with results from recent literature. At the near infrared wavelength of $780\nm$
the experimental single-to-double ionization ratio is reproduced up to intensities of $4\times 10^{14}\Wcm$, 
and two-electron energy spectra and joint angular distributions are presented. 
The time-dependent surface flux (tSurff) approach is extended to full 3+3 spatial dimensions and 
systematic error control is demonstrated. \hide{Estimated errors range from below $5\%$ to $\lesssim 30\%$ 
\commAZ{mention errors here}
for the majority of the results, except for the five-fold differential spectra at infrared wavelength, 
where only qualitative results can be given.}
We analyze our differential spectra in terms of an experimentally accessible quantitative measure of correlation.
\end{abstract}

\maketitle

\section{Introduction}

\tempsubsection{three body coulomb breakup is key}

Understanding the quantum dynamics of the three body Coulomb problem
is a fundamental and ongoing challenge. As one of its most elementary
realizations, the three body breakup of Helium atoms  by short
and intense laser pulses has been examined extensively over the
past few decades. The dynamics and correlation in this simple process 
provide an important model for understanding more complex few-body phenomena. 

\tempsubsection{IR mechanisms and models - we need accurate data to compare to.}

A multitude of mechanisms for the ionization of He have been identified 
in the various intensity and wavelength regimes. For example, at long laser wavelengths
in the near infrared (IR) and at high intensities, double ionization (DI) is well described 
by two independent tunnel ionization events. 
At lower intensities, however, the DI yield predicted by such a model
is far too small.
The observed enhancement of the DI yield by several orders of magnitude
compared to the expectation for a sequence of 
two independent processes
appears not just in Helium~\cite{Walker1994} but in many atoms and molecules~\cite{Jesus2004}.
To explain the discrepancy
various so-called non-sequential double ionization (NSDI) mechanisms have been proposed (an overview can be found in, e.g.,~\cite{Becker2008}).
\tempsubsection{IR - details of the mechanisms}
The NSDI models are all based on the recollision scenario of  the ``three-step-model''~\cite{Corkum1993}, 
where one electron escapes via tunnel ionization, picks up energy from the laser field and is 
redirected back on the ion when the field reverses its direction. Upon recollision
the first electron can interact with the second electron and the nucleus in various ways. 
%\commentVisible{[all acronyms are cited in~\cite{Emmanouilidou2011,Emmanouilidou2011(Taylor)}]}
It may directly dislodge the second electron leading to simultaneous ejection (SE) of both electrons,
also called the direct pathway.
It may also lift the second electron to an excited state from which the field can ionize it on its own,
also called recollision induced excitation with subsequent ionization (RESI), 
which can result in a delayed pathway for the second electron that leads to emission into 
opposite directions~\cite{Haan2006,Shaaran2010}.
The first electron can also form a bound compound with the second electron and the nucleus which survives
for at least a quarter of a cycle (doubly delayed ejection, DDE)~\cite{Emmanouilidou2011}, 
a process which can also occur when the kinetic energy of the 
recolliding electron is smaller than the excitation energy of the ion.
Details and the interplay of all these mechanisms are not yet fully understood,
and require testing against accurate benchmark data.

\tempsubsection{XUV DI - mechanisms}

Another process that has received significant attention is two-photon DI of the He atom.
This occurs in the extreme ultraviolet (XUV) when the energy of two photons suffices
for double ionization and it dominates up to the threshold for single photon double ionization.
The process necessarily 
probes correlation dynamics, most simply by shake-up into an excited ionic state during detachment of the 
first electron, followed by second ionization of the
excited ionic state. Also more direct processes were considered, that involve correlated initial and
final states, see e.g.~\cite{Nikolopoulos2007}.
In numerical simulations, consensus appears to be arising in recent literature
for  photon energies $\omega \approx 40\eV \sim 54\eV$, see e.g.~\cite{Feist2008}.

\tempsubsection{experiments}

At present, the most complete experimental tests for the mechanisms
discussed 
above are provided by detecting, in coincidence, the momenta
of two particles from the fully fragmented state consisting of the
two ionized electrons and the remaining ion. Momentum
imaging techniques like COLTRIMS~\cite{Ullrich2003} and continued advances in laser technologies
have opened up the possibility to study the correlated three-particle
Coulomb breakup in Helium experimentally on its intrinsic time scale,
see for example~\cite{Avaldi2005,Staudte2007,Rudenko2007,Rudenko2008,Fleischer2011}.
Due to the high binding energies of the Helium atom and the resulting low efficiency
of laser induced ionization, many DI experiments are
performed with other targets, such as Neon~\cite{Fleischer2011} or
Argon atoms~\cite{Stevenson2008,Pfeiffer2011,Bergues2012,Henrichs2013,Liu2014a}.
%\commentVisible{
%The measured two electron momentum spectra contain, in principle,
%all information about the involved processes.: actually, time is missing }
Final state correlations
are one of the more accessible observables, but also more convoluted
questions are being investigated, like the exact release time of the
electrons~\cite{Pfeiffer2011}. Although experimentally more challenging,
data for Helium is often preferable as the additional electrons in multi-electron targets
 complicate analysis and model building.

\tempsubsection{numerics: what people have done mostly is XUV}

Apart from the discussion of simplifying models, 
significant effort has been invested into numerical
computations. The main goal is to obtain reliable benchmark data 
that are not obscured by experimental limitations. Advances in numerical 
techniques and ever increasing computational resources allowed for extensive 
\textit{ab initio} calculations of the Helium system. 

At short laser wavelengths, 
where dynamics are
initiated by the absorption of only few photons, fully differential
photo-electron spectra and cross sections have been computed 
using a wide array of methods. 
As laser-atom interaction at XUV wavelength is weak, time-independent perturbative
methods can be applied \cite{Kheifets2006,Horner2007}. For such approaches, 
representation of the double continuum is the major computational task, for which
a variety of techniques exist, see for example the discussions in
\cite{Malegat2010,Argenti2013}. Two-photon double ionization has also been 
successfully calculated using R-matrix Floquet theory~\cite{Feng2003,Burke1991}.
A large number of calculations resorted to numerical solutions 
of the time-dependent Schr\"odinger equation (TDSE) in full dimensionality, see
for example~\cite{Schneider2006(Collins),Feist2008,Palacios2010,Nepstad2010,Zhang2011,Argenti2013,Jiang2013,Liu2014}.

At longer wavelengths such computations remain challenging. The reason
is the inapplicability of perturbative approaches, which only allows
for time-dependent non-perturbative methods relying on numerical solutions
of the TDSE.  Unfortunately, calculations are subjected to a rather unfavorable 
scaling of the problem size with the
wavelength $\lambda$ of the laser pulse proportional to  $\lambda^{p}$ with
$p\gtrsim 7$ (see \ref{sec:ProblemHelium})
\tempsubsection{numerics: what people have done at longer $\lambda$}
As a consequence, already at wavelengths ${\lambda\!\sim\!400\nm}$ full
dimensional numerical computations are scarce~\cite{Parker2003,Parker2006,Emmanouilidou2011(Taylor)}.
Progress at longer wavelengths like the experimentally relevant Ti:sapphire wavelength of
$780\nm$ has only been made recently. Weak infrared (IR) dressing
fields, unable to ionize on their own, were combined with XUV
pulses in~\cite{Liu2014} to study their effect on
joint angular distributions for coplanar emission.
In Ref.~\cite{Hu2013} a study of the enhancement of double emission by an XUV pulse
in presence of a moderately intense single cycle IR pulse 
required over 4000 cores of a Supercomputer for computation.
In Ref.~\cite{Parker2003}, at very high computational expense, ionization due to a pure $780\nm$ IR pulse
could be computed, but no spectra were given.

\tempsubsection{what we present in this paper}

In this article, we present accurate {\it ab initio} DI spectra at both
XUV and IR wavelength. In the XUV regime, we reproduce results from 
literature and corroborate the consensus that has emerged in recent 
publications for observables such as the total DI cross section, two-electron 
energy distributions and triply differential cross sections. Compared to literature,
we were able to significantly reduce the computational effort.

At the near IR wavelength of $780\nm$, we present differential spectra
for intensities up to $4\mycdot 10^{14}\Wcm$ and pulse durations $\gtrsim\!10\fs$.
The measured single-to-double ionization ratio --- the IR double-ionization ``knee'' --- 
is reproduced. We further present two-electron energy 
distributions and up to five-fold differential momentum spectra. 

Throughout, convergence is studied systematically and provides error estimates from
below $5\%$ in the XUV regime to $\lesssim 30\%$ for a large part of the IR data.
At certain energies errors of three-fold differential spectra are beyond these 
values and the five-fold differential spectra at IR wavelength must be considered 
only as qualitative results, as convergence 
could not yet be achieved. For quantitative analysis we suggest a measure of 
correlation that is directly applicable to experimental double-emission spectra. 
 
\tempsubsection{our method}

All calculations are based on an extension of the time-dependent surface flux (tSurff) method
to the six-dimensional double emission problem. 
The three-dimensional single particle version of the method~\cite{Tao2012(Scrinzi)}
has been applied to several systems~\cite{Hofmann2014,Santra2014,Yue2014,Majety2015a,Majety2015gauge,Torlina2015}.
A first formulation for double-emission from a 1+1 dimensional model system was 
given in~\cite{Scrinzi2012}. Here we give the 3+3 dimensional version of tSurff that
is needed for double ionization in realistic systems.

\tempsubsection{description of how paper is organized}
The paper is organized as follows.
In the next section we specify the problem and summarize the general approach.
Then we describe the tSurff method for double emission followed
by a discussion of the method's limitations and error control. 
In section~\ref{sec:numerical-implementation} we present
details of the implementation. Then we compare our results with various
previous theoretical and experimental publications at XUV wavelengths to verify the validity
of the method. Finally, we present our results for 
fully differential DI spectra in the IR regime
in section~\ref{sec:resultsIR}.
Technical details of the method are presented in several appendices.

\section{tSurff for double photo-emission\label{sec:the-problem-and}}

\subsection{Direct computation of double emission\label{sec:ProblemHelium}}

A Helium atom interacting with an external electric field is described
by the Hamilton operator
\begin{equation}
H(t)=H_{\mathrm{ion}}(t)\otimes\id
+\id\otimes H_{\mathrm{ion}}(t)
+\frac{1}{|\vec{r}_{1}-\vec{r}_{2}|}
\label{eq:heliumhamiltonian}
\end{equation}
where 
\begin{equation}\label{eq:Hion}
H_{\mathrm{ion}}(t)=-\frac{\Delta}{2}+\mathrm{i}\vec{A}(t)\cdot\vec{\nabla}-\frac{2}{r}
\end{equation}
is the single electron Hamiltonian of the ionic problem. Atomic units
($\hbar=m_{e}=e^2=4\pi\epsilon_0 \equiv 1$) are used unless indicated otherwise.
The motion of the nucleus is neglected.
%Also, all Coulomb interactions, including those with the electric field
%$\vec{\mathcal{E}}(t)=-\partial_{t}\vec{A}(t)$, are treated classically.
For the interaction with the field $\vec{\mathcal{E}}(t)=-\partial_{t}\vec{A}(t)$ we employ
the dipole approximation, which is appropriate down to wavelengths in the XUV 
%ultra violet wavelengths and above
($\gtrsim10\nm$).
The reasons for choosing the velocity gauge form of the dipole interaction
are detailed in  section~\ref{sec:absorption-gauge}.

Here we discuss only linear polarization and choose the $z$-axis to coincide with the laser 
polarization direction. One reason for this choice is that the recollision mechanism largely
responsible for double ionization is most effective for linear polarization. A more mundane reason 
is that cylindrical symmetry reduces the spatial dimensions to five instead of six for general polarization.

The goal is to extract both
single- and double-ionization photo-electron spectra generated by
an external laser pulse from the solution of the TDSE
\begin{equation}
\mathrm{i}\ddt\Psi(t)=H(t)\Psi(t)\label{eq:TDSE}
\end{equation}
starting from the groundstate.
% Here we are interested
%in linear polarized pulses of the form $\vec{A}(t)=A\vec{e}_{z}\cdot C(t)\cdot\sin(\omega t)$
%with amplitude $A$, envelope $C(t)$ and carrier frequency $\omega$.

The direct approach to this problem consists of two steps, both of
which are numerically challenging. 
First, the multichannel wavefunction $\Psi$ including double continuum
contributions needs to be computed at the end of the pulse, a task whose complexity
depends on the laser parameters. In particular, it scales very unfavorably with 
the laser wavelength due to a simultaneous expansion in momentum, space, and time. 
At high intensities $I$ and long wavelengths $\lambda$, the peak electron momenta 
are dominated by the vector potential $A\!\propto\!\lambda\sqrt{I}$. For correctly 
representing such momenta, thinking in terms of grids, the grid point density must be 
increased proportional to the maximal momentum $p_{\text{max}}$. The increase of both
$p_{\text{max}}\propto\lambda\sqrt{I}$ and pulse duration $\propto \!\lambda$ let the spatial 
extension of the solution grow to a maximal radius $R_{\rm max}\propto \!\lambda^2 \sqrt{I}$. 
As a result, the required number of discretization points in the laser 
direction grows as $\propto\!\lambda^3 I$. Any discretization, not only grids, 
is subjected to the same general scaling. 
Considering only the growth of the radial discretization of the two electrons and assuming
pulse durations $\propto\!\lambda$, the computational effort
for solving a two-electron system grows as $\propto\!\lambda^{7}I^{2}$.
This is a conservative estimate as it ignores the effects on the angular degrees of
freedom and on the time step size.

The second difficulty arises in the analysis of the wavefunction $\Psi$ after the end of 
the pulse. For extracting the double-emission amplitudes
one would need to know the two particle stationary scattering solutions
$\chi_{\vec{k}_{1},\vec{k}_{2}}(\vec{r}_{1},\vec{r}_{2})$ for asymptotic 
outgoing single particle momenta $\vec{k}_{1}$ and $\vk_2$.
Such solutions are not available and analysis involves additional, hard to control
approximations. A strategy for bypassing this problem is to propagate $\Psi$
 to sufficiently long times after the end of the pulse
and then extract the relevant dynamical information entirely from
the asymptotic region where effects of electron repulsion are disregarded. 
In this approximation, the scattering solutions are
products of single-particle scattering wavefunctions 
$\chi_{\vec{k}_{1},\vec{k}_{2}}(\vec{r}_{1},\vec{r}_{2})
\approx\chi_{\vec{k}_{1}}(\vec{r}_{1})\chi_{\vec{k}_{2}}(\vec{r}_{2})$,
where popular choices for $\chi_{\vk_i}$ are Coulomb or
plane waves. The effect of this approximation can be 
controlled by propagating further into the asymptotic region. Various other strategies
for the analysis of the multichannel wavefunction have been proposed,
which all incur some form of inconvenience, ranging from large
computational costs to inability to extract differential information.
Discussions can be found in~\cite{Malegat2010} or~\cite{Argenti2013}
and the references therein.

The direct approach has been implemented by several groups using various
combinations of strategies to tackle both steps. 
A particularly convincing example is Ref.~\cite{Feist2008}, where the 
time-dependent close-coupling scheme (TDCC)~\cite{Pindzola2007} was
implemented in a finite element discrete variable representation (FE-DVR) 
\cite{McCurdy2004(Rescigno),Schneider2002} to compute differential two-photon 
cross sections. Box sizes of up to $800\au$ were
used to propagate up to $21\fs$ after the XUV laser pulse in order
for the projection onto products of energy-normalized Coulomb waves
to be accurate. 
The same numerical methods were used by various other groups
to study DI by few photons~\cite{Pazourek2011,Zhang2011,Jiang2013},
and the effects of an assisting IR field~\cite{Liu2014}.
In~\cite{Nepstad2010} spatial discretization was by B-splines 
and analysis by projection onto products of uncorrelated numerical single-electron
continuum states.
In~\cite{Palacios2010} further propagation after the end of the pulse was avoided.
The wavepacket was analyzed in terms of three-body scattering solutions that are obtained
by exterior complex scaling. This implies the product form $\chi_{\vk_1}\chi_{\vk_2}$ 
only outside the simulation box, where
for the parameters of \cite{Palacios2010} a box size of  $\sim130\au$ was required.
A strategy employing a finite differences discretization and extraction of DI spectra using masks
was used to study DI processes at $390\nm$~\cite{Parker2006}.
In~\cite{Emmanouilidou2011(Taylor)} the same method was used to
analyze the relative importance of various DI pathways at this wavelength,
which required box sizes of up to $1200\au$.
In~\cite{Hu2013} an XUV pulse was used to enhance photo absorption from a very
short IR pulse of moderate intensity. The computations employed similar 
numerical techniques as~\cite{Feist2008} and were conducted
on a grid with over $300\au$ radial extension.

\subsection{Double emission by tSurff\label{sec:tsurffDI}}

In the direct approach, a significant part of the computational effort
goes into reproducing basically trivial dynamics: far away from the
nucleus and when the electrons are far apart, the only relevant interaction
of the electrons is with the external electric field. 
That is the point which is exploited by the tSurff method. 
Numerical simulation is limited to the volume inside a ``tSurff radius'' $R_c$,
where interactions between the charges are important. Beyond that, spectra
can be reconstructed from analytically known solutions for an electron in a dipole field.
The numerical solution is spatially terminated by placing an efficient absorber outside $R_c$.

The following sections convey the basic idea of tSurff and present all formulae needed for 
computing double emission spectra in $3+3$ dimensions. For a more detailed discussion 
the reader is referred to Refs.~\cite{Tao2012(Scrinzi),Scrinzi2012}. In separate subsections we explain 
the gain of the method, its scaling compared to the direct approach, and the special case of single emission 
from two-electron systems.

\subsubsection{tSurff for a single particle problem}

Single electron spectral densities can be expressed through spectral amplitudes $b(\vk)$ 
as
\beq
P(\vk)=|b(\vk)|^2.
\eeq
The fundamental idea of scattering theory is that at sufficiently large times $t\geq T$ and sufficiently large
distances $|\vr|\geq R_c$ the time evolution of the system equals the free time evolution such that the spectral amplitudes
 at momentum $\vk$ reaching a detector outside $R_c$ can be computed as 
\beq\label{eq:tsurff1}
b(\vk)\approx(2\pi)\inv{3/2}\int_{|\vr|>R_c} \!\!\!\!\mathrm{d}\vr\,
\mathrm{e}^{-\mathrm{i}\vk\vr}\Psi(\vr,T).
\eeq
The approximate sign accounts for the fact that exponential tails of the bound states will
extend to infinite distances $|\vr|>R_c$ and that at any finite $T$ some low momentum content
of $\Psi(\vr,T)$ will not have left the region $|\vr|<R_c$. Both errors rapidly decrease with 
growing $T$ and $R_c$.

The analysis  of $\Psi(\vr,T)$ in terms of field-free scattering solutions is only 
meaningful in absence of any external field. For laser-ionization we must choose $T$ after the end of the laser pulse. 
As typical laser pulse durations are several hundred atomic units, $\Psi(\vr,T)$ in general extends over
large distances. The overwhelming part of this extension is brought about by essentially free motion
at distances $|\vr|>R_c$. In tSurff, this motion is not simulated numerically, but will be inferred
from analytically known solutions.

For the practical implementation of the idea, let $H(\vr,t)$ be the Hamiltonian of the system in the field
and let $H_V(\vr,t)$ be a time-dependent Hamiltonian with
\beq\label{eq:Hc}
H(\vr,t)=H_V(\vr,t),\quad |\vr|>R_c,
\eeq
for which we know solutions with the desired asymptotic momenta $\vk$
\beq
\mathrm{i}\ddt \chi_\vk(\vr,t) = H_V(\vr,t)  \chi_\vk(\vr,t).
\eeq
Here we have in mind the Hamiltonian of a free electron in a dipole field
\beq\label{eq:Hvolkov}
H_V=-\frac{\Delta}{2} + \mathrm{i}\vA(t)\cdot\vna
\eeq
with the Volkov solutions
\beq
\chi_{\vec{k}}(\vec{r},t) = (2\pi)^{-3/2}\mathrm{e}^{\mathrm{i}\vec{k}\vec{r}}\mathrm{e}^{-\mathrm{i}\Phi(\vec{k},t)}
\eeq
and the Volkov phase
$\Phi(\vec{k},t)=\int^{t}\!\mathrm{d}\tau\,\big(\vec{k}^{2}/2-\vec{k}\vec{A}(\tau)\big)$.
Introducing the notation
\begin{equation}
\Theta(\vr):=\begin{cases}
0 & \text{for }|\vec{r}|<R_{c}\\
1 & \text{else}
\end{cases}\label{eq:thetafunction}
\end{equation}
we can write the spectral amplitude (\ref{eq:tsurff1}) for our photo-emission problem as
\beq
b(\vk)\approx \l \chi_\vk(T)|\Theta|\Psi(T)\r,
\eeq
where we have dropped the Volkov phase factor $\exp[\mathrm{i}\Phi(\vk,t)]$ 
at $t=T$.
By taking the time derivative and 
integrating over time we find
\beq\label{eq:tsurffInt}
\l \chi_\vk(T)|\Theta|\Psi(T)\r=
\mathrm{i}\int_{-\infty}^T \!\!\mathrm{d}t\, 
\l \chi_\vk(t)|H_V\Theta-\Theta H|\Psi(t)\r.
\eeq
The lower integration boundary of $-\infty$ is to be understood as any time before the onset of the
laser field.
By Eq.~(\ref{eq:Hc}) the difference of operators reduces to a commutator
\beq
H_V\Theta-\Theta H=[H_V,\Theta]=
[-\frac{\Delta}{2}+\mathrm{i}\vA(t)\cdot\vna,\Theta],
\eeq
which is a flux operator on the surface $|\vr|=R_c$ with a time-dependent correction accounting for the
action of the external field, see also appendix~\ref{sec:appen:commutator}.
For evaluating the tSurff integral (\ref{eq:tsurffInt}) we no longer need
to integrate over the long range behavior of $\Psi(\vr,T)$. Instead we have a time integral over the surface flux.
The time-dependent correction in the surface flux operator accounts for the acceleration of the electron by the 
dipole field outside $R_c$ and ensures that the flux is counted into the correct final momentum.

An important complication of this idea is caused by the long-range nature of the Coulomb potential.
It is well known that for scattering potentials with Coulomb-like asymptotics $V(r)\sim 1/r$, 
standard scattering theory suffers from a divergent error in the scattering phases at any finite $T$ and $R_c$.  
In time-independent scattering this can be remedied by replacing the plane wave in Eq.~(\ref{eq:tsurff1}) 
with a scattering solution of the corresponding 
Coulomb problem. For tSurff this is not viable, as no analytical solutions are known, where both 
the laser and the asymptotic Coulomb field are taken into account. 

We solve this problem pragmatically by using
$R_c$ as a convergence parameter, i.e.\ increasing it until further changes fall below a desired accuracy limit.
In practice this is achieved by multiplying the potentials by a function $f_{\al,\be}(r)=1$ for $r<\al$, $=0$ for $r>\be$.
For the transition from $\al$ to $\be$ we employ a third order polynomial
such that the derivatives are continuous at $\al$ and $\be$. 
We usually choose $\be=R_c$.

When highly accurate results are needed the strategy above becomes costly and defeats the original purpose of tSurff, i.e.\
to keep $R_c$ small. For such situations, analytic corrections to the plane wave solutions, such as the Eikonal-Volkov 
solutions introduced in \cite{Smirnova2008} may be helpful. 

\subsubsection{tSurff in 3 + 3 spatial dimensions}

The tSurff method was extended to two-particle emission
in~\cite{Scrinzi2012} for the example of two one-dimensional particles. 
The straight-forward extension for two three-dimensional particles is given here. 

The experimentally observed momentum distribution density is expressed by the two-electron 
spectral amplitudes as
\begin{equation}
P(\vec{k}_{1},\vec{k}_{2})=|b(\vec{k}_{1},\vec{k}_{2})|^{2}.\label{eq:spec}
\end{equation}
For the computation of the $b(\vec{k}_{1},\vec{k}_{2})$, we truncate the
long range tails of the nuclear Coulomb potentials as in
the single electron case.
In three-body breakup there is the additional complication that
motion never can be considered as free near $\vr_1=\vr_2$,
where the two electrons repel each other irrespective of their distance from 
the nucleus. This difficulty is characteristic for the asymptotics
of many body scattering. We avoid it 
by suppressing the repulsion for $r_1,r_2\geq R_c\geq \be$:
\begin{equation}
\frac{1}{|\vr_1-\vr_2|}\mapsto\frac{ f_{\al,\be}(r_{1})f_{\al,\be}(r_{2})}{|\vr_1-\vr_2|}.\label{eq:truncationinterval}
\end{equation}
This kind of  approximation is not specific to tSurff. Any asymptotic analysis where the two-particle scattering 
functions are approximated as a product of single-particle functions implies that interparticle interactions are 
neglected in the asymptotic region. This is the case for all direct methods discussed above. Here we make this 
approximation manifest by suppressing the electron repulsion outside $R_c$: rather than having a built-in
error in the asymptotic analysis, we make a consistent spectral analysis of the approximate system.

We would like to point out that the approximation may possibly be avoided in tSurff \cite{Scrinzi2012}: an exact solution for two 
electrons in a laser field can be given in relative and center-of-mass coordinates $(\vr_1+\vr_2)/2,\vr_1-\vr_2$ if nuclear potentials 
can be neglected. In practice, this involves rather complicated transformations
of the surfaces which we have not attempted to implement.

\begin{figure}[H]
\begin{center}
\includegraphics[width=4cm]{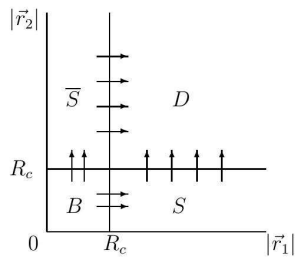}
\end{center}
 \caption{Spatial partitioning of the wavefunction: at large times, regions $B$, $S$ and $\overline{S}$, and 
$D$ correspond to bound, singly ionized, and doubly ionized parts, respectively. 
Arrows symbolize the flux across the region boundaries. \label{fig:regions}
}
\end{figure}

We divide the space
of the radial coordinates of the two electrons into four regions (Fig.~\ref{fig:regions}):
$B:=[0,R_{c}]\times[0,R_{c}]$, $S:=[R_{c},\infty)\times[0,R_{c}]$,
$\overline{S}:=[0,R_{c}]\times[R_{c},\infty)$ and $D:=[R_{c},\infty)\times[R_{c},\infty)$.
Analogous to the single-electron case, for large $R_{c}$ and at large $T$, the
wavefunctions in these regions approximately contain the bound, singly
ionized, and doubly ionized parts of the wavefunction, respectively. 

With this partitioning, similar as for a single particle, 
we obtain the double scattering amplitude by analyzing the solution
in region $D$ only:
\beq\label{eq:bDouble}
b(\vec{k}_{1},\vec{k}_{2})\approx\LL\chi_{\vec{k}_{1}}(T)\chi_{\vk_2}(T)|\Theta_1\Theta_2|\Psi(T)\rr,
\eeq
with $\Theta_i:=\Theta(\vr_i)$, $i=1,2$, see Eq.~(\ref{eq:thetafunction}).
The double brackets $\LL \ldots \rr$ emphasize that integration is over both spatial coordinates. 
The bra function is the product of two Volkov waves $\chi_{\vk_i}(\vr_i,T)$, $i=1,2$, 
with the respective 
asymptotic momenta. As all interactions have been switched off beyond $R_c$, these are correct asymptotic solutions
for the truncated problem: in region $D$ the truncated Hamiltonian is
\beq
\left.H(t)\right|_D=H_{V}(t)\otimes\id+\id\otimes H_{V}(t),\label{eq:hamiltonOnD}
\eeq
with the Volkov Hamiltonian $H_V(t)$, Eq.~(\ref{eq:Hvolkov}), on the respective coordinates.

By similar steps as for a single particle, 
the spectral amplitude~(\ref{eq:bDouble}) can be evaluated 
as the time integral over fluxes $F$  and $\overline{F}$ from regions $S$ and $\overline{S}$ into $D$
\begin{equation}
b(\vec{k}_{1},\vec{k}_{2},T)=
\int_{-\infty}^{T}\!\mathrm{d}t\,
\big(\underbrace{F(\vec{k}_{1},\vec{k}_{2},t)}_{S\to D}+\underbrace{\overline{F}(\vec{k}_{1},\vec{k}_{2},t)}_{\overline{S}\to D}\big).
\label{eq:bAsFluxIntoD}
\end{equation}
By exchange symmetry, we have $F(\vk_1,\vk_2)=\overline{F}(\vk_2,\vk_1)$.
Details of the derivation can be found in~\cite{Scrinzi2012}.

The flux $F$ is determined by the time evolution on $S$,
where the Hamiltonian is given by
\begin{equation}
\left.H(t)\right|_S=H_{V}(t)\otimes\id+\id\otimes H_{\mathrm{ion}}(t)\label{eq:HonS}
\end{equation}
with $H_{\mathrm{ion}}$ obtained from Eq.~(\ref{eq:Hion}) by truncating the Coulomb tails. 
Solutions for $H(t)|_S$ can be written as the product of a Volkov wave $\chi_{\vk_1}(\vr_1,t)$ and
the solution of an ionic problem $\varphi_{\vk_1}(\vr_2,t)$ on $|\vr|<R_c$. With this the 
flux from $S$ into $D$ is given as 
\begin{equation}
F(\vec{k}_{1},\vec{k}_{2},t)=
\langle\chi_{\vec{k}_{2}}(t)|[H_{V},\Theta_{2}]|\varphi_{\vk_1}(t)\rangle,
\label{eq:Fiswahat}
\end{equation}
where we again refer to \cite{Scrinzi2012} for the derivation.
For details on the evaluation of $F$, see appendix~\ref{sec:appen:commutator}.

It is important to note that the ionic factor $\varphi_{\vk_1}(\vr_2,t)$ for $\vr_2$ depends on the 
asymptotic momentum $\vk_1$ in $\vr_1$-direction.  
This coupling occurs, as $\varphi_{\vk_1}$ is determined by an {\em inhomogeneous} TDSE, 
where the inhomogeneity accounts for the flux from region $B$ into $S$, 
see Fig.~\ref{fig:regions}. The flux $B\to S$ is correlated, such that each momentum component
$\vk_1$ contributes differently to the wavefunction in $\vr_2$ direction.  
The inhomogeneous equation for $\varphi_{\vk_1}(\vr_2,t)$ is
\begin{equation}
\mathrm{i}\ddt\varphi_{\vk_1}(\vr_2,t)=
H_{\rm{ion}}(t)\varphi_{\vk_1}(\vr_2,t)-C_{\vk_1}(\vr_2,t),
\label{eq:TDSEsOnS}
\end{equation}
with the source term
\begin{equation}
C_{\vec{k}_{1}}(\vr_2,t)=
\int \!\mathrm{d}\vr_1\, 
\overline{\chi_{\vec{k}_{1}}(\vr_1,t)}
\big[H_{V}(\vr_1),\Theta_{1}\big]
\Psi(\vr_1,\vr_2,t)
\label{eq:sourceterm}
\end{equation}
and initial condition
\beq
\varphi_{\vk_1}(\vr_2,t=-\infty)=0.
\eeq
For evaluating (\ref{eq:sourceterm}) we need the values and
derivatives of $\Psi(\vec{r}_{1},\vr_2,t)$ on the boundary $|\vr_1|=R_c$
between domains $B$ and $S$. This requires the solution of the full 6-dimensional 
(in case of linearly polarized pulses 5-dimensional) two-electron TDSE 
on $B$. Beyond $R_{c}$ the two particle wavefunction $\Psi$
can be disposed off by absorption. Details on the choice of absorber
and its implementation are given in section~\ref{sec:absorption-gauge}.

\subsubsection{Procedure and Computational Gain\label{sec:computationalgain}}

In principle and in practice the computation of the surface flux
and the computation of spectra are separate steps. 

For the surface flux, one solves the TDSE on the region $B$ and absorbs the solution outside.
Absorption can be done at low computational cost and  without any distortions.
During propagation values and derivatives of $\Psi(\vr_1,\vr_2,t)$ at $|\vr_1|=R_c$
are stored for a sufficiently dense time grid $t_n:=n\Delta t$. 
The storage intervals are determined by the maximal energy $E_{\rm max}$ that one wishes to resolve.
The fastest relevant phase oscillations have the period $2\pi\hbar/E_{\rm max}$.
One period needs to be sampled at about 8 time points resulting in $\Delta t\approx \pi\hbar/(4E_{\rm max})$.
Even for the largest of our applications, file sizes remained within 
reasonable limits of $\sim \!\!100\,\mathrm{GB}$.

In the second step, using the surfaces stored on file, the integrals (\ref{eq:bDouble}) are 
computed for the desired 6-dimensional grid of momentum pairs $\vk_1,\vk_2$. 
The sheer amount of information available in double ionization spectra
makes this an inherently large computation. 
We use the same grid of $\vk\up{\mu}_1=\vk\up{\mu}_2$, $\mu=1,\ldots,K$ for both momenta. 
The $\vk\up{\mu}_i$ can be conveniently chosen on a quadrature grid in polar coordinates, which allows for 
easy transition to a representation in terms of partial waves. 

The scattering amplitudes (\ref{eq:bAsFluxIntoD}) are accumulated in the following steps:
\begin{itemize}
\item[(a)] For given $\vk_1\up{\mu}$, solve for $\varphi_{\vk_1\up{\mu}}(\vr_2,t)$, Eq.~(\ref{eq:TDSEsOnS}).
\item[(b)] At $t=t_n$, get $F(\vk_1\up{\mu},\vk_2\up{\nu},t_n)$ $\forall\vk_2\up{\nu}$, Eq.~(\ref{eq:Fiswahat}).
\item[(c)] Add $F(\ldots)/\Delta t$ into $b(\vk_1\up{\mu},\vk_2\up{\nu})$, Eq.~(\ref{eq:bAsFluxIntoD}).
\end{itemize}

The main computational gain of the method is that the complete, correlated 
two-electron problem needs to be solved only in the confined domain $B$. 
There is no relevant overhead in computing and storing the surface flux.

At long wavelengths and correspondingly low photon energies $\hbar\om$ the radial extension required for domain $B$ is approximately 
given by the quiver amplitude $Q:= A_{\rm max}/\om$ of the electron in a field with peak vector potential $A_{\rm max}$
(see sections~\ref{sec:limitationsControl} and \ref{sec:DIrecollision}). 
In contrast, using the standard approach for computing differential DI spectra, the radial extension of the 
computation box is given by the maximal distance $R_\mathrm{max}$ 
that the wavepacket can travel until the end of time propagation.
This distance is roughly given by the maximal momentum $p_\mathrm{max}$
times the duration $2\pi n/\om$ of an $n$-cycle pulse.
For long wavelengths, photo-electron energies extend to 10 times the ponderomotive energy $U_p=A_{\rm max}^2/4$.
This corresponds to maximal momentum $p_\mathrm{max} = \sqrt{2 E_{\mathrm{max}}} \approx \sqrt{5}A_{\rm max}$.
Considering both radial directions, at IR wavelength the tSurff volume  $B$ is reduced by a  factor
\begin{equation}
\left(\frac{R_\mathrm{max}}{Q}\right)^2 \approx \left(\frac{2\pi n\sqrt{5} A_{\rm max}/\om }{A_{\rm max}/\om}\right)^2 \approx 200\, n^2,
\end{equation}
comparing to  any approach where the full wavefunction $\Psi(\vr_1,\vr_2,T)$ is 
required. In particular, we remark that $B$ is independent of pulse duration $n$, allowing for long pulses with only linear
increase in computational cost. 

This significant gain for the two-electron computation comes at the expense of the additional  steps~(a)-(c). 
Their actual cost depends on the desired momentum range and resolution.
The main computational effort is to solve the inhomogeneous ionic TDSE for each $\vk_1\up{\mu}$, step~(a). 
Note that, contrary to the full two-electron problem, the equations for different $\vk_1$ are not coupled and lend themselves to 
straight forward parallelization. In addition, the effort for one ionic problem 
scales as $\lambda^4I$ as compared to the $\lambda^7I^2$ scaling of the two-electron problem,
compare section~\ref{sec:ProblemHelium}. 
One has to consider that also the detail of information contained in the momentum spectra grows
as $\la^3I$ and the overall scaling of the problem persists, if one endeavors to retrieve
the spectra on the full 6-dimensional momentum grid. This is mitigated by the possibility
to perform the spectral analysis, steps~(a)-(c) only on the subsections of the $(\vk_1,\vk_2)$-space
that are of interest. In the more challenging calculations at IR
 wavelength of $\la=780\nm$, correlated 
time propagation and spectral analysis took about equal times.

\subsubsection{Single ionization spectra}
For completeness we show how single ionization spectra are computed.
The function $\chi_{\vk_1}(\vr_1,T)\varphi_{\vk_1}(\vr_2,T)$ is the component of $\Psi(\vr_1,\vr_2,T)$
on $S$ with asymptotic momentum $\vk_1$. The spectral amplitude of a given ionic channel corresponding to the
state $\phi\up{I}(\vr_2)$, $||\phi\up{I}||=1$, is 
\beq
b\up{I}(\vk_1)\approx\l\phi\up{I}| \varphi_{\vk_1}(T)\r.
\eeq
For the spectral densities, a factor 2 accounts for the symmetric contributions from $S$ and $\overline{S}$
\beq
P\up{I}(\vk_1)=2|b\up{I}(\vk_1)|^2.
\eeq

If double ionization is negligible, the problem can be further simplified by admitting only 
contributions from the source~(\ref{eq:sourceterm}) that will end up in 
channel $I$. We introduce
the ionic solution
\beq\label{eq:channelTDSE}
\mathrm{i}\ddt \phi\up{I}(\vr_2,t) = H_{\rm ion}(t)\phi\up{I}(\vr_2,t)
\eeq
with a {\em final} condition at $t=T$ 
\beq
\phi\up{I}(\vr_2,T)=\phi\up{I}(\vr_2).
\eeq
With this, the channel source is
\beq
C_{\vk_1}\up{I}(\vr_2,t)=\phi\up{I}(\vr_2,t)\l \phi\up{I}(t)|C_{\vk_1}(t)\r
\eeq
and $\varphi_{\vk_1}(\vr_2,t)\propto \phi\up{I}(\vr_2,t)$ fulfills the homogeneous part of Eq.~(\ref{eq:TDSEsOnS}) by definition. One readily sees that
\beq
b\up{I}(\vk_1,t)=\int_{-\infty}^T \!\mathrm{d}t\, \l \phi\up{I}(t)|C_{\vk_1}(t)\r.
\eeq
With all surface values for all times available on disk, one can perform the 
integral starting from $T$ and solve (\ref{eq:channelTDSE}) backward in time.

The non-ionizing condition will be noted, as also in backward propagation the 
$\phi\up{I}(\vr_2,t)$ must remain confined in $[0,R_c]$. Violation of the condition
appears as reflections or, in presence of an absorber, by exponential divergence of 
the solution, rather than exponential damping.
  
The simplification is significant, as the ionic solution $\phi\up{I}(\vr_2,t)$ needs to 
be computed only once for all $\vk_1$. This approach was used in~\cite{Majety2015a} 
and~\cite{Zielinski2015} to compute single ionization spectra from multi-electron 
systems including core polarization and
doubly excited states.

\subsection{Control of convergence\label{sec:limitationsControl}}

\tempsubsection{we need to converge the new parameter $R_{c}$}

Eq.~(\ref{eq:tsurff1}) becomes exact only in the limit $T\to\infty$ and $R_c\to\infty$.
The effects of truncation at finite $T$ and $R_c$ must be examined for 
any given physical situation. This holds not only for tSurff, but any method that 
cannot draw on exact scattering information from other sources. 
In tSurff, the computational
cost increases only linearly with $T$ and therefore convergence
with respect to $T$ is usually easily achieved. Convergence with the radial extension $R_{c}$
of domain $B$ represents a much more difficult task. In the result sections 
\ref{sec:resultsUV} and~\ref{sec:resultsIR} we will
therefore always demonstrate convergence behavior with the tSurff
radius $R_{c}$, which strongly depends on the observable of interest.

\subsubsection{Single-electron convergence}

The two particle version of the tSurff method inherits
the convergence properties
of its single particle predecessor~\cite{Tao2012(Scrinzi)}.

\tempsubsection{low energy parts want long time propagation}

First, contributions to the photo-electron spectrum are only taken
into account from those parts of the wavefunction which passed through
the tSurff surface at $R_{c}$ at time $T$, see 
equations~(\ref{eq:tsurffInt}) and~(\ref{eq:bAsFluxIntoD}).
In order to capture low energetic contributions correctly, time propagation
must continue until some time after the end of the pulse. This limitation
applies whenever a finite domain of the wavepacket $\Psi(T)$ is excluded from analysis. 
All approaches listed above, except for~\cite{Palacios2010}, are affected by this.

\tempsubsection{truncation of core potential implies errors}

Second, the truncation of the nuclear potential beyond $R_{c}$ transforms
the Coulomb potential into a short range potential,
where the missing long range behavior modifies electron
trajectories. This is particularly important near thresholds
with strong modification of the Rydberg states and low energy scattering.
The nuclear potential truncation is the most severe
approximation introduced by tSurff in the single particle case, and
this carries over to the two particle situation. At the moment, the
effects can only be controlled by increasing the truncation radius
$R_{c}$. It may also be possible to introduce corrections to the
scattering wavefunction $\chi_{\vec{k}}(\vec{r},t)$ and thereby
dispose of the nuclear potential truncation completely, which is a
goal for future work. We are currently investigating the possibility
to replace the pure Volkov waves with Eikonal Volkov waves \cite{Smirnova2008}.

\tempsubsection{rydberg state averaging wants long time propagation}

Third, highly excited bound states can extend well beyond $R_c$,
which conflicts with the assumption that any wavefunction part 
beyond $R_c$
belongs to the continuous spectrum.
The effect of a persistent finite probability density at $R_c$
are oscillations of the spectral density with respect to energy and $T$.
Increasing $R_{c}$ can reduce these artifacts as the bound state density 
decays exponentially. However, for highly excited states this decay is slow 
and moving to large $R_c$ comes at large computational
cost. If the bound states are known well enough, they can be projected out
once the pulse is over, see~\cite{Tao2012(Scrinzi)}. A generally applicable method
to suppress these oscillations is to average the spectral amplitudes over a range of propagation times $T$,
which has proven to be simple and efficient, with quick convergence and little extra
computational cost. Details are given in appendix~\ref{sec:appen:averaging}.

\tempsubsection{absorption radius must conserve recollisions}

Fourth, a practical lower limit for the size of the simulation box 
is set by the quiver radius $Q$ of the motion of a free electron in the laser 
field: as electrons may return from distance $Q$ 
into regions of strong interactions to cause, for example, recollisions, the dynamics over 
the whole region must be included. The irECS absorber 
used for our computations does, in principle, maintain
the full dynamics, but it can do so only if discretized with a large number of coefficients, 
see~\cite{Scrinzi2010} and section~\ref{sec:absorption-gauge}.
Rather than placing $R_c$ inside the quiver radius with a generous number
of discretization points for irECS, one better moves $R_c$ to near $Q$ with only a few points for irECS.
In this way one also benefits from a better approximation of the long range Coulomb interaction.

\subsubsection{Two-electron convergence}

Specific for two-electron systems is the error introduced
by approximating the asymptotics of the exact
scattering solution in product form 
$\chi_{\vk_1,\vk_2}\approx\chi_{\vk_1}\chi_{\vk_2}$.
This error decreases with growing $R_c$. As the essence of tSurff is to keep $R_c$ as small
as possible, it is affected most acutely by this. In the direct approach,
the product ansatz is made typically only beyond $|\vr|\gtrsim 100$, either 
explicitly or implicitly as in Ref.~\cite{Palacios2010}. We will show that,
depending on the observable in question, tSurff radii as small as $R_c\approx 20$ 
can give sensible results.
Some observables are strongly affected by this approximation:
whenever ``postcollision'' interaction, i.e.\ repulsion
between electrons far from the nucleus, is important,
a product description is bound to fail.
This is most pronounced for side-by-side
double emission, where the two electrons are in close proximity for
long times. The relevant distances depend on the details of the process~\cite{Feist2009}. 
If these distances lie beyond practical $R_c$ values, tSurff would need to be
amended by fully including post-collision electron-electron
interaction~\cite{Scrinzi2012}, but such an approach has not been proven yet in practice.

\section{Numerical Implementation\label{sec:numerical-implementation}}

\subsection{Basis}

The tSurff method requires the numerical solution of the full two
particle TDSE (\ref{eq:TDSE}) on domain $B$ and several single particle
TDSEs (\ref{eq:TDSEsOnS}) on domain $S$. In this section we present
our choice for the discretization, discuss its merits, and compare
it to other popular choices.

\tempsubsection{angular basis on B}

We expand the two-electron wavefunction on $B$ into single particle
spherical harmonics $Y_l^m(\Omega)\equiv Y_l^m(\theta,\varphi)$
\begin{align}
&\Psi(\vec{r}_{1},\vec{r}_{2},t)|_B=\label{eq:discretization} \nonumber \\
&\sum_{l_{1}m_{1}}\sum_{l_{2}m_{2}}Y_{l_{1}}^{m_{1}}(\Omega_{1})Y_{l_{2}}^{m_{2}}(\Omega_{2})R_{l_{1}m_{1}l_{2}m_{2}}(r_{1},r_{2},t).
\end{align}
For linearly polarized pulses, the magnetic quantum number $M$ is conserved and $m_{1}=M-m_{2}$.
The sums are truncated when convergence is reached, an example for 
the truncation of the $l_{1}$-$l_{2}$-grid at IR wavelengths is given in section~\ref{sec:convergenceOfDisc}.
% The two-electron wavefunction on $S$ factorizes into the analytically known Volkov solution 
% $\chi_{\vk_1}(\vr_1,t)$ and a solution in $\vr_2$-direction, that needs to be determined
% numerically:
% \beq
% \Psi(\vec{r}_{1},\vec{r}_{2},t)|_S
% =\int\mathrm{d}\vec{k}_{1}\,\chi_{\vec{k}_{1}}(\vec{r}_{1},t)\varphi_{\vec{k}_{1}}(\vec{r}_{2},t).
% \eeq
The numerical single particle functions $\varphi_{\vec{k}_{1}}(\vec{r}_{2})$ needed in region $S$  (Eq.~(\ref{eq:TDSEsOnS}))
are also expanded into spherical harmonics:
\begin{equation}
\varphi_{\vec{k}_1}(\vec{r}_2,t)=
\sum_{l_2m_2}Y_{l_2}^{m_2}(\Omega_2)R_{l_2m_2}(r_2,t)
\label{eq:RaddiscOnS}.
\end{equation}

\tempsubsection{radial discretization}

The radial functions are discretized by a finite element method. 
For $R_{l_1m_1l_2m_2}$ the $r_1r_2$-plane is divided into rectangular patches 
$[r_1\up{n_1-1},r_1\up{n_1}]\times[r_2\up{n_2-1},r_2\up{n_2}]$ and we write
\beq
R_{l_1m_1l_2m_2}(r_1,r_2)=\sum_{n_1,n_2=1}^N R_{l_1m_1l_2m_2}\up{n_1,n_2}(r_1,r_2,t),
\eeq
where 
\bea
\lefteqn{ R_{l_1m_1l_2m_2}\up{n_1,n_2}(r_1,r_2,t)=}\nonumber
\\&&\sum_{p_1,p_2} f\up{n_1}_{p_1}(r_1)f\up{n_2}_{p_2}(r_2) c\up{n_1,n_2}_{p_1l_1m_1p_2l_2m_2}(t)
 \label{eq:radialProductR}
\eea
is expanded into products of one-dimensional finite element basis functions 
$f\up{n_1}_{p_1}(r_1)$ and $f\up{n_2}_{p_2}(r_2)$. The $f\up{n}_{p}(r)$
are high order polynomials that are confined to the interval $[r\up{n-1},r\up{n}]$.
The last interval is let to extend to $r\up{N}=\infty$ with an exponential factor $\exp(-\al r)$ with $\al\lesssim 1$.
We usually choose $R_c=r\up{N-1}$ and absorption acts only on the last interval, see section~\ref{sec:absorption-gauge}.
An analogous expansion into $f_{p_2}\up{n_2}$ is used for the single particle radial functions $R_{l_2m_2}$.
This type of radial discretization was introduced in~\cite{Scrinzi2010}, 
also see appendix~\ref{sub:Radial-Basis:-Continuity}.

Denoting the expansion coefficients for $\varphi_{\vk_1}(\vr_2,t)$ as the components $d_j$ of a vector $\vd$ 
with the single-electron multi-index $j:=(n_2,p_2,l_2,m_2)$, the time evolution of $\vd$ is
\beq
\mathrm{i}\ddt \vd(t) = 
\mS_2\inv1\left[\mH_{\rm ion}(t)\vd(t) - \vC(t)\right],
\label{eq:discretizedTDSEsOnS}
\eeq 
where $\mH_{\rm ion}$ and $\mS_2$ are the ionic Hamiltonian and overlap matrices 
with respect to the single electron indices $j$. 
$\vC$ is the vector  of overlaps $\l j | C\r$ of the source
term $C(\vr_2,t)$, Eq.~(\ref{eq:sourceterm}),
with the basis functions $|j\r$.
In region $B$, the equation of motion for the expansion coefficients $c_i$
with the two-electron multi-index   $i:=(n_1,p_1,l_1,m_1,n_2,p_2,l_2,m_2)$
is
\begin{equation}
\mathrm{i}\ddt\vec{c}(t)=\mS^{-1}\mH(t)\vec{c}(t)\label{eq:discretizedTDSE},
\end{equation}
where $\mH(t)$ and $\mS$ denote Hamiltonian
and overlap matrices \wrt $i$.
As the Hamiltonian is local, i.e.\ it contains only multiplication and differential operators,
the localization of the radial basis on the $n_1n_2$-patches 
produces block structured matrices $\mH$, $\mS$ and $\mH_{\rm ion}$. 
Note that the overlap matrix is not the unit matrix $\mS_{ij}\neq\delta_{ij}$,
as the radial basis functions $\{f_k\up{n}\}$ are not orthonormal.

The ansatz function~(\ref{eq:radialProductR}) is not guaranteed to be
continuous across the element  boundaries $r_i\up{n}$. Constraining $\vc$
to ensure continuity effectively connects the separate matrix blocks. As a result,
one has to solve a linear system of the form $\vb=\mS^{-1}\va$ 
at each time step. Considering the very large basis size, this
may appear a daunting task. Closer inspection shows that the inverse
has tensor product form:
\beq
\mS\inv1=\mS_1\inv1\otimes\mS_2\inv1.
\eeq
Further, the overlap matrices $\mS_1$ and $\mS_2$ can be reduced to near diagonal form,
with only two non-zero elements off diagonal for every element boundary point $r_1\up{n_1}$ and $r_2\up{n_2}$. 
The exact inverses of such matrices can be computed as a diagonal
with a low rank correction such that the floating operations count 
becomes negligible. Details are given in
appendix~\ref{sub:Radial-Basis:-Continuity}.

\subsection{Time propagation}

For the time integration of equations~(\ref{eq:discretizedTDSE})
and~(\ref{eq:discretizedTDSEsOnS}) 
we use the standard fourth order Runge-Kutta (RK4)
algorithm with adaptive step size control. Step size control is important
as at long wavelengths, depending on intensity, the time evolution can be 
driven by the external field and strongly 
vary with the strength of the vector potential $|\vA(t)|$. 
Experiments with exponential integrators such as Arnoldi
and Magnus propagators appear to indicate that these play out their advantages mostly over intervals 
where the time evolution operator is well approximated by a time-constant matrix exponential. 
For strong near IR fields, the gain in time step size  
can not outweigh the overhead of these more complex methods. 

The rather small, but cheap time steps of the RK4 are no major disadvantage for tSurff,
as we want to sample the solution at time intervals $\Delta t$ 
that are short for physical reasons, see section \ref{sec:computationalgain}. 
In fact, we found that the time step returned by the RK4 step size control
was typically only factors $5-10$ smaller than the required sampling intervals 
for resolving the full energy range. 

This said, it may well be that the use of, for example, higher order implicit methods
or methods for time-dependent matrix exponentiation \cite{Trotter1959,suzuki1985} 
would improve performance.

\subsection{Absorption and choice of gauge\label{sec:absorption-gauge}}

\tempsubsection{any absorber is ok. we suggest irECS}

For solving the two particle TDSE (\ref{eq:discretizedTDSE}) and
the single particle TDSEs (\ref{eq:discretizedTDSEsOnS}) we
want to start absorption right at $R_c$ to minimize the computational effort.
For correct results, the absorber must be perfect, as any reflections
immediately corrupt the surface values.
We employ infinite range exterior complex scaling (irECS)~\cite{Scrinzi2010} for this purpose.
Exterior complex scaling (ECS) is an analytical continuation method and has the useful property
to preserve (in principle) the full information of the dynamics even
in the absorbing region.
This allows for particle re-entry from the scaled into the unscaled region,
although in numerical computations excessive excursion into the scaled region
will lead to accumulation of numerical errors.
Typically, small tails of rescattering wavepackets moderately extending into the absorbing region 
are sufficiently undisturbed, which allows for box sizes $R_c$ close to or even below the quiver amplitude $Q$.
Numerical evidence for this fact was given in~\cite{Scrinzi2010}.
irECS is a particular form of discretizing ECS that gives perfect absorption with about $20$ discretization
coefficients on each radial coordinate.

\tempsubsection{choice of gauge}

Note that exterior complex scaling only works with suitable interactions.
It can be used in the velocity gauge representation of the dipole operator $\mathrm{i}\vec{A}(t)\cdot\vec{\nabla}$,
but not in length gauge $\vec{E}(t)\cdot\vec{r}$.
Fortunately, velocity gauge is also numerically favored for 
strong field problems~\cite{Cormier1996,Majety2015gauge}.

\tempsubsection{implementation with finite elements}

Implementation of irECS in a finite element scheme is straight forward~\cite{Scrinzi2010},
details can be found in appendix~\ref{sec:appen:absorption}.
In time-independent situations, standard ECS has been applied also with other discretizations, 
including FE-DVR~\cite{McCurdy2004(Rescigno)}. 
Severe problems of accuracy and stability had been reported for ECS when applied to the TDSE
using FE-DVR discretizations~\cite{Tao2009}. Only recently we observed that the
reported errors are not related to ECS and that standard ECS as well as irECS can be applied in finite difference and 
FE-DVR schemes with comparable efficiency as in the present finite-element implementation \cite{weinmueller2015}.

\subsection{Electron-electron interaction\label{sec:e-e-int}}

When solving the TDSE (\ref{eq:discretizedTDSE}) on domain $B$,
the electron-electron interaction represents the major computational
challenge. It is the only part of the Hamilton operator that does not 
factorize into tensor products with respect to the two particles.
Using the multipole expansion we can express
the matrix connecting the radial $(n_1,n_2)$ patch with $(n_1',n_2')$ as
\bea
\lefteqn{\langle\Psi^{(n_1'n_2')}_{l'_1 m'_1 l'_2 m'_2}|\frac{1}{|\vec{r}_{1}-\vec{r}_{2}|}|\Psi^{(n_1n_2)}_{l_1 m_1 l_2 m_2}\rangle\nonumber}
% \\& = &
% \sum_{\lambda}\frac{4\pi}{2\lambda+1} G_{l_{1}',\lambda,l_{1}}^{-m_1',m_1'-m_1,m_1}G_{l_{2}',\lambda,l_{2}}^{-m_2',m_2'-m_2,m_2}\nonumber
\\& = &
\sum_{\la\mu}\frac{4\pi}{2\lambda+1} \l Y^{m'_1}_{l_1'}Y_\la^\mu|Y_{l_1}^{m_1}\r \l Y^{m'_2}_{l_2'}|Y_\la^\mu Y_{l_2}^{m_2}\r\nonumber
\\&& 
\times
\underbrace{\l R\up{n_1'n_2'}_{l_1'm_1'l_2'm_2'}|
\frac{\min(r_1,r_2)^\la}{\max(r_1,r_2)^{\la+1}}|R\up{n_1n_2}_{l_1m_1l_2m_2}\r}_{=:\mV\up\la}.
\label{eq:multipole}
\eea
The electron-electron interaction matrix is full with respect to all indices. 
For finite angular expansion, the sum over $\la,\mu$ remains finite and, in many cases,
can be truncated at relatively low $\la$ without introducing relevant numerical error. 
The operations count for matrix-vector application scales  $\propto P^{4}$
for a radial expansion of maximal degree $P-1$. This is larger than for all 
other terms in the Hamiltonian, which have tensor product structure with operations 
count $\propto P^{3}$.

However, as pointed out in~\cite{McCurdy2004(Rescigno)},
in a polynomial basis the scaling can be reduced to $P^3$.
In appendix~\ref{sec:app:multipole} it is shown that 
the radial multipole matrices $\mV\up\la$, Eq.~(\ref{eq:multipole}),
can be exactly represented by a multiplication on an $R$-point
quadrature grid ($R:=2P-1$) that is independent of $\la$:
\beq
\mV\up{\la}=\left(\mT\up{n_1}\otimes\mT\up{n_2}\right)^\mathrm{T} 
\mD\up\la 
\left(\mT\up{n_1}\otimes\mT\up{n_2}\right)\label{eq:e-e-int-diagonal},
\eeq
where he transformation matrices $\mT\up{n_i}$ are $R\!\times\! P$
and $\mD\up\la$ is a {\em diagonal} $R^2\!\times\!R^2$ matrix.
Applying (\ref{eq:e-e-int-diagonal}) to a radial coefficient vector of length $P^2$ 
has the operations count $2PR(R+P)+R^2$. In practice, one can admit a minor 
quadrature error and reduce the quadrature grid to $R=P$ points without compromising any
of the results reported below. Be cautioned that $\mD\up\la$ is not just
${\min(r_1,r_2)^\la}/{\max(r_1,r_2)^{\la+1}}$ evaluated on the quadrature grid points. Its correct form together with 
other details of the procedure are given in 
appendix~\ref{sec:app:multipole}.

\subsection{Computational resources}
At XUV wavelength tSurff computations can be performed on the work station scale. 
For example,  on a 16 core shared memory machine the computation of the two-photon double 
ionization cross section (see section~\ref{sec:two-photon}) took about 12 hours with $R_c=30\au$ and 
about 60 hours with $R_c=80\au$ for each photon energy.

Resource requirements are not significantly increased by the addition 
of a weak IR field as in section~\ref{sec:xuv-ir}: 
with a total of $551$ partial waves on a 16 core machine, 
the largest computation with $R_c=25$  ran for a maximum of two days.

Computations at strong IR fields are the most challenging.
Yet, the largest of the computations reported in section~\ref{sec:DIrecollision} 
used 128 cores, running for 10 days in an MPI parallelized scheme  
for solving the full two-electron problem on domain $B$, see section~\ref{sec:computationalgain}.
Additional 5 days were required to compute the solutions on domain $S$ for a 
$\vec{k}_1$-$\vec{k}_2$-grid dense enough for extraction of the fully differential data.

This should be put into relation to the much larger computing facilities 
employed in the direct approaches.  For example, Ref.~\cite{Hu2013}, used 4000 cores
for computations including only $295$ partial waves at a weak IR field (no run times are quoted). Resource consumption 
was not reported in
Refs.~\cite{Parker2006,Feist2008,Feist2009,Pazourek2011,Palacios2010}, but in all cases computations
were performed at supercomputing centers using large scale machines.

\subsection{Other implementation options}

Several other strategies for the discretization and time propagation
of the TDSE have been applied in the literature. 
In~\cite{Feist2008,Pazourek2011,Hu2013}
the time-dependent close-coupling scheme~\cite{Pindzola2007} and
FE-DVR~\cite{McCurdy2004(Rescigno),Schneider2002} were used. In~\cite{Parker2003,Parker2006,Emmanouilidou2011(Taylor)}
finite differences were used for discretization, in~\cite{Nepstad2010}
B-splines. 

For time propagation, more sophisticated procedures
were chosen than the present RK4:  the short
iterative Lanczos method~\cite{Smyth1998} in~\cite{Feist2008},  the
real-space-product algorithm~\cite{Collins1998} in~\cite{Hu2013}, or a
 Crank\textendash{}Nicolson method in~\cite{McCurdy2004(Rescigno)}.
For the time propagation we keep the simple choice for the reasons discussed above.

\tempsubsection{On the Issue of FE-DVR}
We chose full finite elements as a discretization for the present calculations, as tSurff had been 
proven to function in this framework. 
The compatibility of the tSurff method with other discretizations
has been investigated only recently \cite{weinmueller2015}.
In view of these recent results, FE-DVR appears to bear the potential for further dramatic
reduction of operations count and improve scalability of tSurff on parallel machines.

There are several other optimizations one can think of, like 
low rank description of the electron-electron interaction at large distances,
a non-square spatial domain for $B$, or alternative time propagators.
Investigation of these technical options is, however, not the purpose of this paper.

\section{Observables}

Here we introduce the various physical quantities which will be examined
in the following result sections.
All observables are derived from the fully, five-fold differential photo-electron spectrum,
which we will refer to as the two-electron probability density 
\begin{equation}
P(E_{1},E_{2},\Omega_{1},\Omega_{2})=k_1k_2P(\vk_1,\vk_2)
\end{equation}
For the linearly polarized pulses considered here, the spectra are independent of the sum of the azimuthal angles $\varphi_1+\varphi_2$.
%and which contains the entire observable information of 
%the DI scattering process.
The $E_{i}=k_i^2/2$ are the final kinetic energies and $\Omega_i:= (\theta_i,\varphi_i)$
are the emission angles.
%Visualizing this 6-dimensional quantity in an informative way is a challenge.
The total DI yield is given by
\begin{equation}
Y:=\int\hspace{-5bp}\int\hspace{-5bp}\int\hspace{-5bp}\int\!\mathrm{d}E_{1}\mathrm{d}E_{2}\mathrm{d}\Omega_{1}\mathrm{d}\Omega_{2}\, P(E_{1},E_{2},\Omega_{1},\Omega_{2}).\label{eq:yield}
\end{equation}
In the regime of multi-photon perturbation theory, one can define
the total $N$-photon cross section as
\begin{equation}
\sigma_{N}:=\frac{\omega^{N}Y}{\int\!\mathrm{d}t\, I(t)^{N}},\label{eq:nphotoncrosssection}
\end{equation}
where $I(t)$ is the laser intensity profile and $\omega$ is the photon
energy (see~\cite{Feist2008,Zhang2011} and references therein). 
Another popular quantity is the triply differential cross section (TDCS) defined as
\begingroup\makeatletter\def\f@size{9}\check@mathfonts\def\maketag@@@#1{\hbox{\m@th\normalsize\normalfont#1}}
\begin{equation}
\frac{\mathrm{d}\sigma_{N}
}{\mathrm{d}E_{1}\mathrm{d}\Omega_{1}\mathrm{d}\Omega_{2}}:=\frac{\omega^{N}}{\int\!\mathrm{d}t\, I(t)^{N}}\int\!\mathrm{d}E_{2}\, P(E_{1},E_{2},\Omega_{1},\Omega_{2})\label{eq:TDCS}
\end{equation}
\endgroup
which 
is usually evaluated for coplanar geometry, $\varphi_{1}-\varphi_{2}=0$ or $=\pi$.

This observable is experimentally accessible as a nuclear recoil momentum distribution,
as summarized in~\cite{Zhang2011}.
These cross sections take into account
the Fourier width of the pulse by the energy integrations. Therefore,
as long as the photon energy is defined sharp enough such that no alternate
reaction channels open up they are in good approximation independent
of the exact pulse shape.

In contrast, the energy probability distribution, given by
\begin{equation}
P(E_{1},E_{2}):=\int\hspace{-5bp}\int\!\mathrm{d}\Omega_{1}\mathrm{d}\Omega_{2}\, P(E_{1},E_{2},\Omega_{1},\Omega_{2})\label{eq:energyprobdist}
\end{equation}
sensitively depends on the exact pulse shape, as was noted in \cite{Feist2008}  and will be shown below. 
The same holds true
for the coplanar joint angular distribution (JAD) at fixed energies $E_{1}$ and $E_{2}$,
\begin{equation}
P_{E_{1},E_{2}}(\theta_{1},\theta_{2}):=P(E_{1},E_{2},\Omega_{1},\Omega_{2})\Big\vert_{\varphi_{1}=\varphi_2}.
\label{eq:jad}
\end{equation}
Here it is convenient to consider $P_{E_{1},E_{2}}(\theta_{1},\theta_{2})$ on $[0,2\pi]\times[0,2\pi]$, 
where $\th_1=\th_2$ is emission into the same direction, ``side-by-side'', and $\th_1=\th_2+\pi$ 
is emission into opposite directions, ``back-to-back''.
Note, that JADs as defined in~\cite{Zhang2011,Liu2014} included
an energy integral, as is appropriate for studying one or two photon
DI. In the non-perturbative regime, where neither $N$-photon
cross sections nor triply differential cross sections are meaningful,
direct evaluation of the differential probability density $P(E_{1},E_{2},\Omega_{1},\Omega_{2})$ is more adequate.

\subsection{Quantifying spectral correlation}
We introduce a measure for angular correlation by a principal
component analysis. Sampling the JAD at fixed energies $E_1,E_2$
on angular grids $\th_1\up{\mu}$ and $\th_2\up{\nu}$, $\mu,\nu=1,\ldots \cA$
we obtain an $\cA\times\cA$ matrix 
\beq
\mP_{\mu,\nu}=P_{E_{1},E_{2}}(\theta_{1}\up{\mu},\theta_{2}\up{\nu}).
\eeq
A singular value decomposition reveals to which extent $\mP$
can be factorized into a product of two (or a few) independent 
single particle distributions on the $\th_1\up{\mu}$ and $\th_2\up{\nu}$ coordinates:
\beq
\mP_{\mu,\nu}=\sum_{\al=1}^\cA s_\al  p^\al_\mu q^\al_\nu.
\eeq
If only a single term contributes to the sum, there is no correlation between the 
two angles, and the more terms are needed, the more correlation we will assign to
the emission pattern. We normalize single particle distributions as $\sum_\mu (p_\mu^\al)^2=\sum_\nu (q_\nu^\al)^2=1$,
and define a measure for the ``length'' of the sum over $\al$ as (compare, e.g.~\cite{Grobe1994})
\begin{equation}
C:=\frac{(\sum_\al s_\al)^2}{\sum_\al s_\al^{2}}.
\label{eq:correlationmeasure}
\end{equation}
For a single non-zero $s_\al$ there is no correlation and $C=1$,
for constant $s_\al$, $C$ is maximal.
For sufficiently dense sampling, $C$ is independent of $\cA$. 

We would like to point out that this measure of correlation is 
readily applicable to experimental data.  
In this way the discussion whether processes occur with strong or
little correlation can be put to a direct experimental test,
independent of the analysis presented in this paper.

\section{Double ionization at XUV wavelengths\label{sec:resultsUV}}

\tempsubsection{goal of section}

We first present results at short wavelengths. At low computational effort we reproduce 
a large body of results available in literature and confirm simple theoretical expectations.

\tempsubsection{convergence parameters vs tsurff parameters}

Our method requires 
the numerical solution of equations~(\ref{eq:discretizedTDSE}) 
and~(\ref{eq:discretizedTDSEsOnS}),
whose convergence depend most notably on the number of partial waves included, Eq.~(\ref{eq:discretization}),
and on the total number of radial coefficients in the finite element scheme, Eq.~(\ref{eq:radialProductR}). 
For short wavelengths, the demand on those discretization parameters is moderate and
a multitude of publications with numerical solutions of the TDSE exist.
All our results presented in this section are converged to $1\%$ or better 
with respect to angular and radial expansion. The variational upper bound for the 
Helium groundstate energy is always below $-2.902\au$, 
which compares well to the exact ground state energy of $E_0=-2.9037\au$. 

\tempsubsection{used parameters}

Propagation times are  $T\geq 8\fs$ after the end of the pulse,
which gives ample time to remove artifacts of Rydberg states by averaging over $T$, 
see section~\ref{sec:limitationsControl}.
The dominant convergence parameter for this section is the tSurff radius $R_c$,
for which we demonstrate convergence of the short wavelength computations explicitly. 
As described in section~\ref{sec:tsurffDI}, in the present implementation of 
tSurff the Coulomb potentials are truncated at $R_c$, 
Eq.~(\ref{eq:truncationinterval}).
In the present section truncation is  over an interval of fixed length $4\au$ 
with smoothing function  $f_{[R_c-4, R_c]}$.

Unless indicated otherwise, results in this section were computed with the pulse
shape used in~\cite{Feist2008}, which allows for direct comparison:
\begin{equation}
A(t)\propto\cos^{2}\Big(\frac{\omega t}{2n}\Big)\sin(\omega t)\label{eq:cos2pulse}
\end{equation}
for times $-n\tau/2\leq t\leq n\tau/2$, where $\omega=2\pi/\tau$ is the 
carrier frequency 
%photon energy
and $n\tau$ is the total duration of the pulse.
The $\cos^2$-envelope is, however, not ideal for emulating realistic experimental pulses as
the spectral decomposition contains spurious side bands, see for example~\cite{Nepstad2010}.
These produce artifacts in single ionization spectra 
and also in DI energy probability distributions.
We demonstrate below that with a pulse envelope that better resembles a Gaussian, such as $\cos^8$, 
the artifacts disappear.

\subsection{Two-photon double ionization cross section\label{sec:two-photon}}

Exposing a Helium atom to a laser field with photon energies $\hbar\omega$ larger than half the double
ionization threshold ($1.45\au$) leads to two photon DI.
At photon energies below the second ionization threshold $\hbar\om<2\au$ double ionization 
necessarily involves electron correlations.
In this regime full agreement among the numerous theoretical approaches
\cite{Bonitz2014,Malegat2012,Palacios2010,Nepstad2010,Feist2008,
Horner2007,Nikolopoulos2007,Ivanov2007,Foumouo2006,Hu2005,Piraux2003,Feng2003,Laulan2003}
has not yet been achieved, not even for the fully integrated
total two photon DI cross section $\sigma_{N=2}$, 
Eq.~(\ref{eq:nphotoncrosssection}).

\begin{figure}[H]
\includegraphics[width=1\columnwidth]{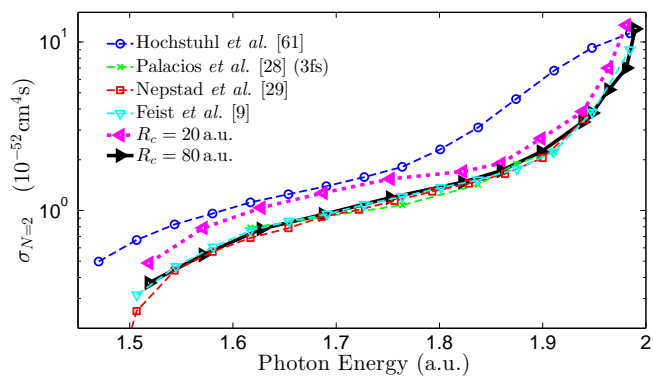}
\caption{\label{fig:TPDIcrosssection_all}
Total two photon
DI cross section as a function of the photon energy.
Results with $R_c=80\au$ and pulse duration $4\fs$ and peak intensity $I=10^{12}\Wcm$ (solid line) 
agree with Refs.~\cite{Feist2008,Palacios2010,Nepstad2010} (dashed lines). 
 Already with $R_c=20\au$ (dotted line) good qualitative agreement is reached.
}
\end{figure}

Fig.~\ref{fig:TPDIcrosssection_all} shows a selection of recent results where approximate
agreement emerges. 
In~\cite{Bonitz2014} the time-dependent full configuration interaction method
was applied using pulses with a bandwidth of $\approx\!0.15\au$.
The authors attribute the large deviation from most other calculations to their method of extracting DI spectra
as well as to the large spectral width of their pulse.
Far from threshold there is good agreement among Refs.~\cite{Feist2008,Palacios2010,Nepstad2010,Piraux2003,Laulan2003}
and with our calculations for $R_c\geq30$, but already at $R_c=20$ we obtain qualitatively correct results.

The divergence of the results \cite{Bonitz2014,Feist2008,Palacios2010,Horner2007} 
is largest near the threshold $\hbar\om\lesssim 2\au$, where the numerical distinction 
between low energy sequential processes and correlated double emission becomes blurred.
Clearly, in this regime results also depend on the spectral width of the pulse.  
At $R_c=80$ we estimate our convergence error to be $\lesssim5\%$
and find agreement with Refs.~\cite{Feist2008,Palacios2010,Nepstad2010}. 
Note that the longest pulse duration of 
\cite{Palacios2010} is $3\fs$ rather than the $4\fs$ of Ref.~\cite{Feist2008}. We verified that the
larger spectral width changes the ratio by less than  3\% for the data point $\hbar\om=53\eV=1.95\au$, which
is the value closest to threshold in \cite{Palacios2010}. The data closest to threshold are found in 
Ref.~\cite{Feist2008}, where we also agree.

\subsection{Energy probability distribution\label{sec:UVenergyDist}}

\begin{figure}[H]
\includegraphics[width=0.5\columnwidth]{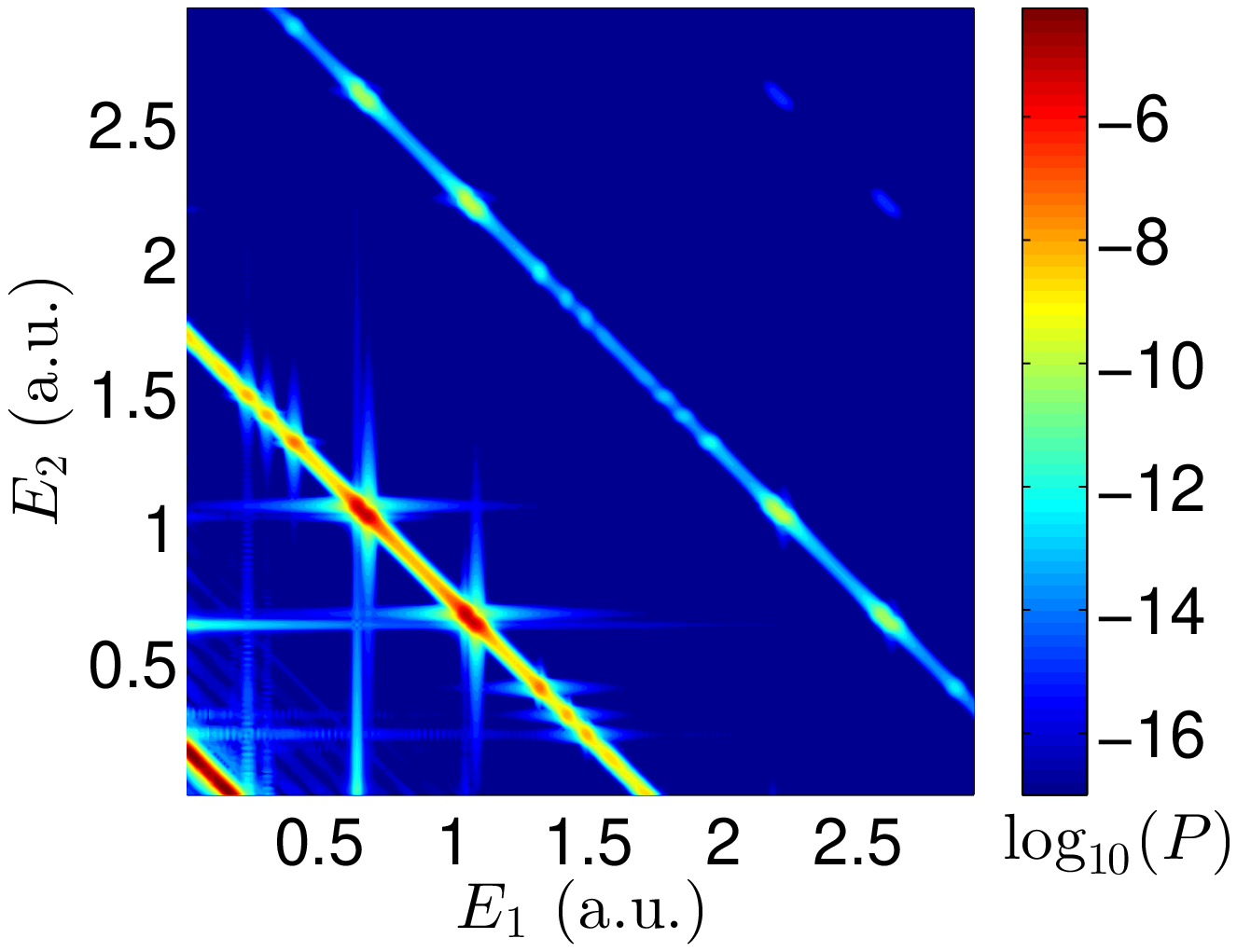}\includegraphics[width=0.5\columnwidth]{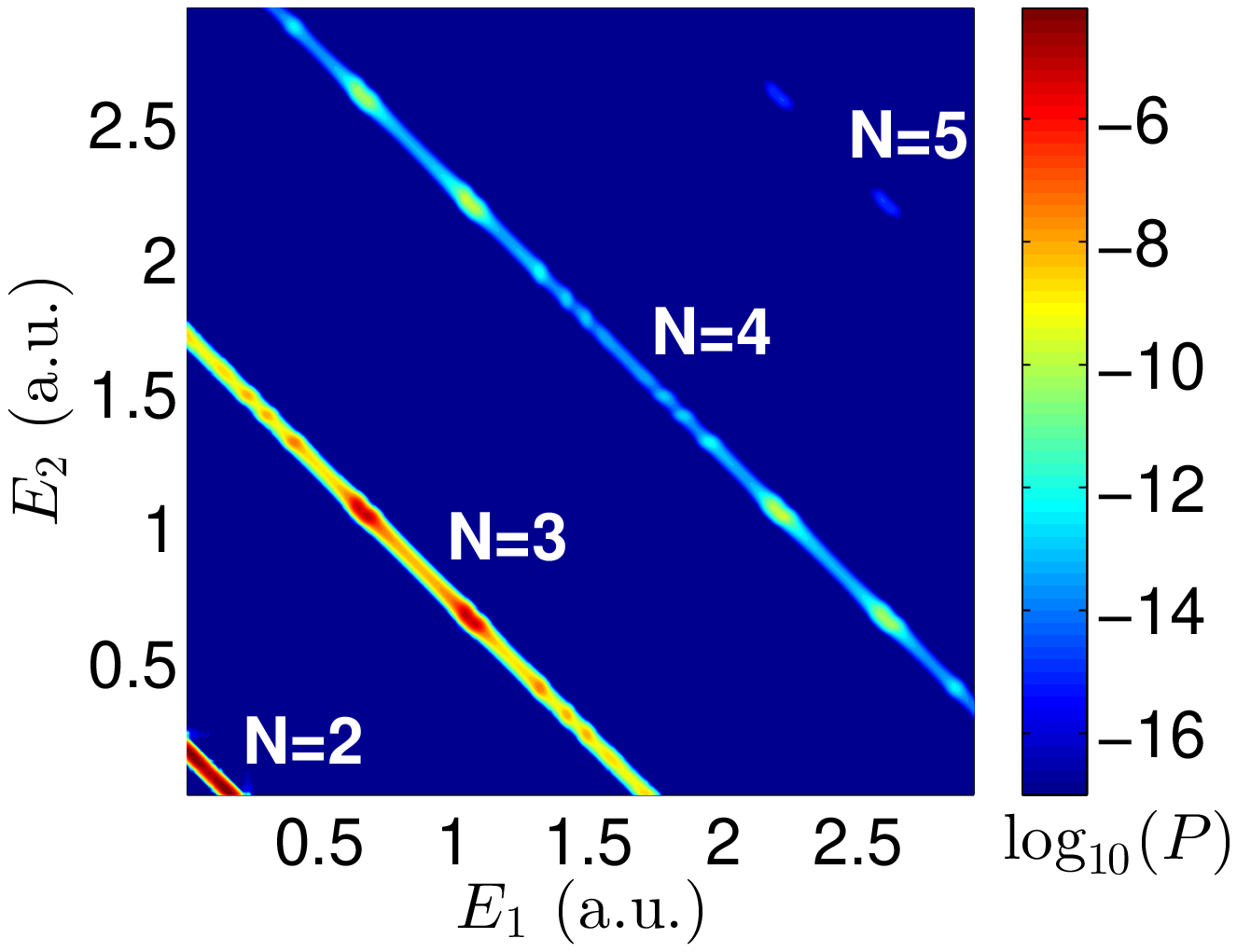}
\caption{\label{fig:XUVenergyDist}
Left: Energy probability distribution $P(E_{1},E_{2})$ for a $n=120$
cycle pulse with photon energy $\hbar\omega=42\eV$ and peak intensity
$I=10^{13}\Wcm$ for a $\cos^2$ pulse envelope.
Right: Calculation with a $\cos^8$ envelope of the same FWHM. Structures generated by the $\cos^2$ sidebands disappear.
}
\end{figure}

Inspecting the angle-integrated energy probability distributions $P(E_1,E_2)$
one sees pronounced two-electron multi-photon lines  (c.f.\ Fig.~\ref{fig:XUVenergyDist})
where the electrons share the total energy  $E_1+E_2=E_0+N\cdot\hbar\omega$, $N=2,3\dots$,
of  $N$ absorbed photons.
The two photon cross section %discussed in the previous section
$\sigma_2$ is 
the integral over the $N=2$ shared energy line.
Along the $N\geq 3$ shared energy lines local maxima 
are found, which is a signature of sequential, uncorrelated double emission.
This is the case if one electron overcomes the first ionization
potential $I_{p}^{(1)}\approx0.9\au$ ending 
up with energy $E_1=-I_{p}^{(1)}+n\hbar\omega$, $n=1,2\dots$,
and in an separate step the second electron gets detached from the ion
by the absorption of two or more photons with final
energy $E_2=-I_{p}^{(2)}+m\hbar\omega$, $m=2,3\dots$.
This process was dubbed ``double ionization above threshold ionization'' (DI-ATI)~\cite{Parker2001}.
Other local maxima along the shared energy lines involve
intermediate excited ionic states, see section~\ref{sec:UVcorrs}.

All these features can
be seen in the energy probability distribution, shown in Fig.~\ref{fig:XUVenergyDist} for the photon energies
$\hbar\omega=42\eV\sim 1.54\au$:
the $N=2$ shared energy line at $E_1+E_2\approx 0.2\au$ does not have particular structure, while
the $N=3$ line at $E_1+E_2\approx 1.7\au$ shows 
pronounced sequential peaks at $(E_{1},E_{2})\approx(0.6,1.1)\au$ and $\approx(1.1,0.6)\au$ and similar at $N=4$.

Fig.~\ref{fig:XUVenergyDist} also demonstrates the effect of the pulse envelope by replacing 
the  $\cos^{2}$ envelope~(\ref{eq:cos2pulse}) by a $\cos^{8}$ 
one with the same full width at half maximum.
The $\cos^{2}$-envelope produces extra DI structures that can hardly be considered as physical.
Such artifacts were already observed in~\cite{Parker2001} but their origin was not linked to envelope effects.
Both computations in Fig.~\ref{fig:XUVenergyDist} used $R_c=20\au$ and $T=1\ps$. 
By the long time propagation any artifacts from Rydberg 
states are safely suppressed (section~\ref{sec:limitationsControl} and appendix~\ref{sec:appen:averaging}).

%\178345 178699

\subsection{Angular distributions}

\tempsubsection{TDCS}

\begin{figure}[H]
\includegraphics[width=1\columnwidth]{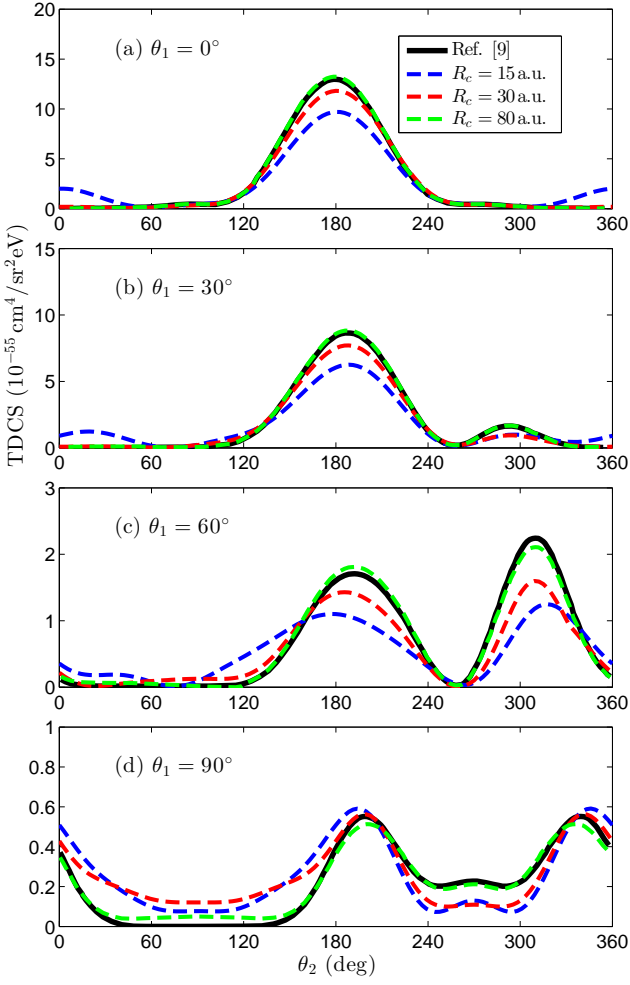}
\caption{\label{fig:TPDI_TDCS}
Triply differential cross section for two
photon DI of He at photon energy  $\hbar\omega=42\eV$ and equal energy sharing 
for different tSurff radii $R_c$.
The same pulse parameters as in Ref.~\cite{Feist2008} were used.
}%figNSDIcompare2.m figNSDIcompare7.m
\end{figure}

The TDCS, Eq.~(\ref{eq:TDCS}), was calculated for $N=2$ at $E_1=0.092\au =2.5\eV$ 
where contributions of equal energy sharing $E_1\approx E_2$ dominate. 
The TDCS, as most angle and energy resolved quantities, is rather sensitive to $R_c$ as
it is strongly affected by postcollision interactions~\cite{Feist2009}. 
In Fig.~\ref{fig:TPDI_TDCS} it can be seen that for equal energy sharing,
even with a box size of $R_c=80\au$, there remain minor quantitative discrepancies with Ref.~\cite{Feist2008}.
The zero in the cross section for side-by-side emission ($\theta_1=\theta_2$)
is reproduced if electron-electron repulsion can act also far from the nucleus.
One can directly see that electron repulsion rather than total box size is responsible, 
by performing computations with $R_c=80\au$ but suppressing electron-electron for all $r_1,r_2>30\au$,
which reproduces the $R_c=30\au$ results.

\tempsubsection{ JADs}

\begin{figure}[H]
\includegraphics[width=0.5\columnwidth]{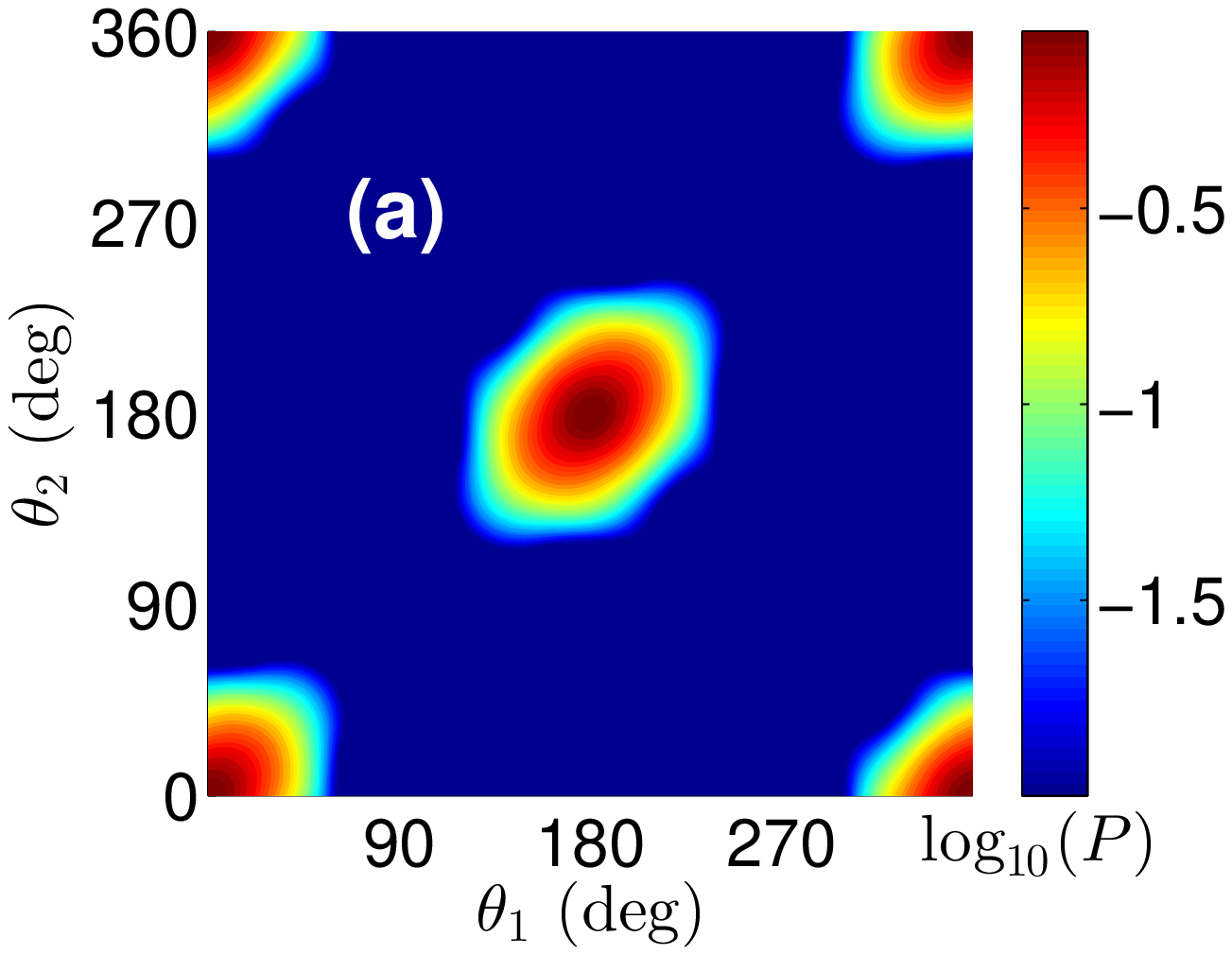}\includegraphics[width=0.5\columnwidth]{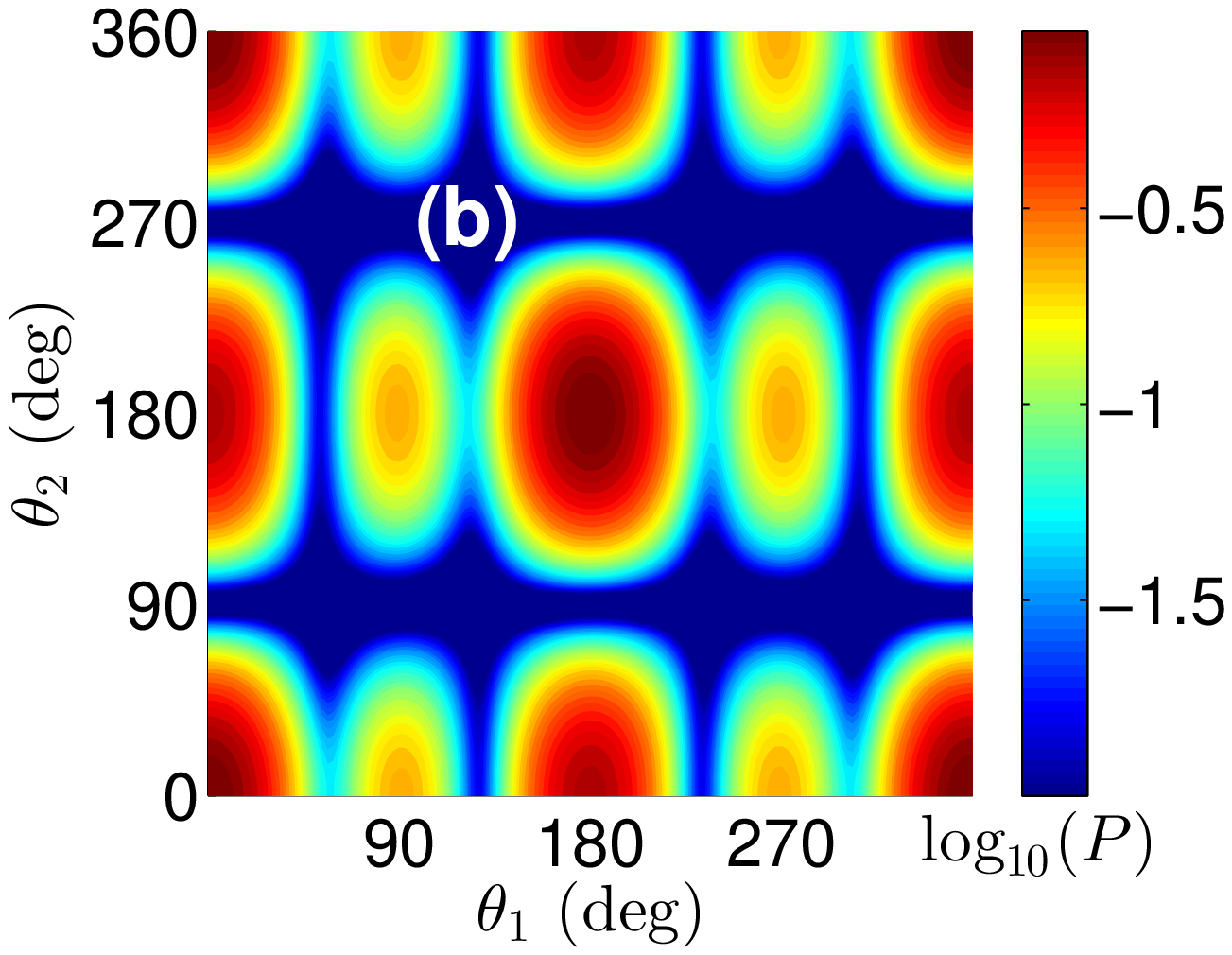}
\caption{\label{fig:XUVjads}
Left: correlated side-by-side emission at three-photon equal energy sharing.
Right: uncorrelated three-photon sequential ionization
at $(E_{1},E_{2})\approx(1.09,0.64)\au$.
 Plots are normalized to $\max_{E_1,E_2}[P(E_1,E_2)]=1$.
}
\end{figure}

The coplanar JADs~(\ref{eq:jad}) provide a two-dimensional differential view on the  
cross sections~(\ref{eq:TDCS}),
 which reveals a pronounced energy dependence of the emission 
patterns. In Fig.~\ref{fig:XUVjads} two exemplary
JADs are shown. At the equal energy sharing point $(E_{1},E_{2})\approx(0.86,0.86)\au$
for three absorbed photons, we observe side-by-side $\th_1\approx\th_2$
emission, see Fig.~\ref{fig:XUVjads}(a). Back-to-back emission is suppressed
due to selection rule C stated in~\cite{Maulbetsch1995}: as the three-photon photo-electron 
states are odd, they have a node at $\vk_1=-\vk_2$.

To contrast this, we picked as a second point the sequentially accessible energies $(E_1,E_2)\approx(1.09,0.64)\au$,
Fig.~\ref{fig:XUVjads}(b),
where emission is almost completely uncorrelated (see also next section) and
well described by the simple angular distribution 
$P(\theta_{1},\theta_{2})\sim\big\vert Y_{2}^{0}(\theta_{1})Y_{1}^{0}(\theta_{2})\big\vert^{2}$.
Qualitatively these structures are already reproduced with box sizes as small as $R_c=15\au$.

\subsection{Angular correlation\label{sec:UVcorrs}}

We already classified, by visual inspection, the angular distributions shown in Fig.~\ref{fig:XUVjads}
as correlated or uncorrelated. For a more quantitative description we use
the correlation measure defined in Eq.~(\ref{eq:correlationmeasure}).

\begin{figure}[H]
\includegraphics[width=1\columnwidth]{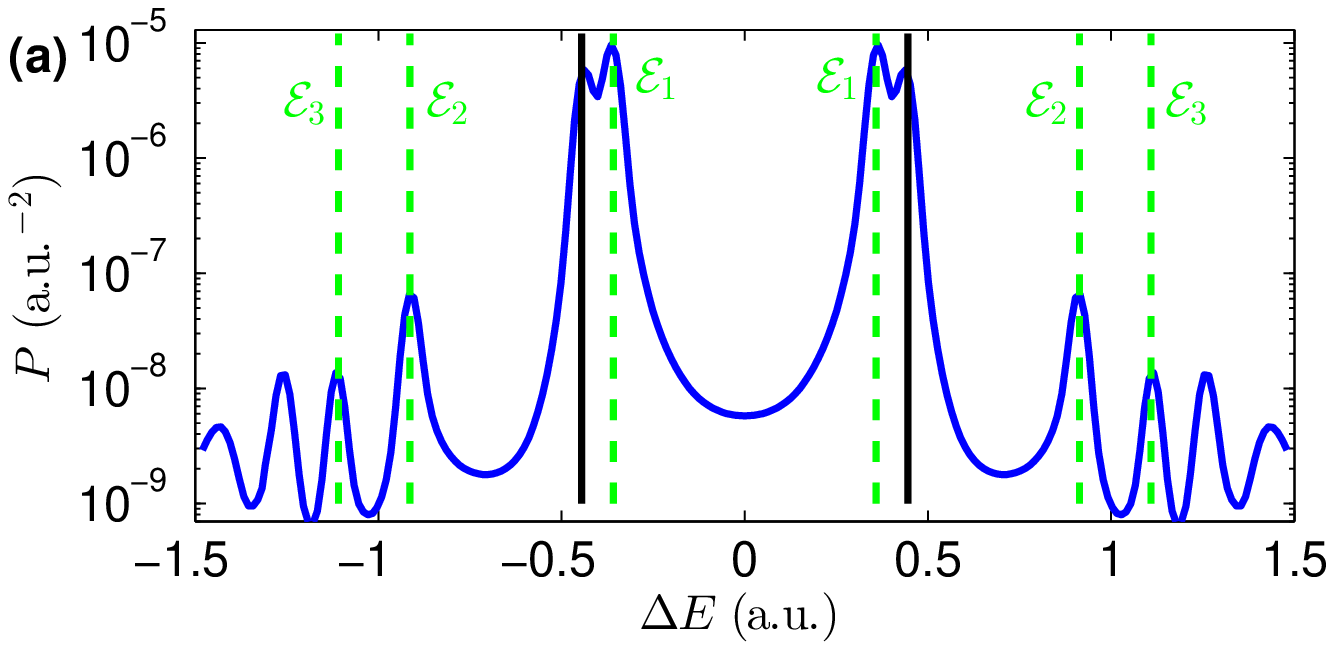}\\
\includegraphics[width=1\columnwidth]{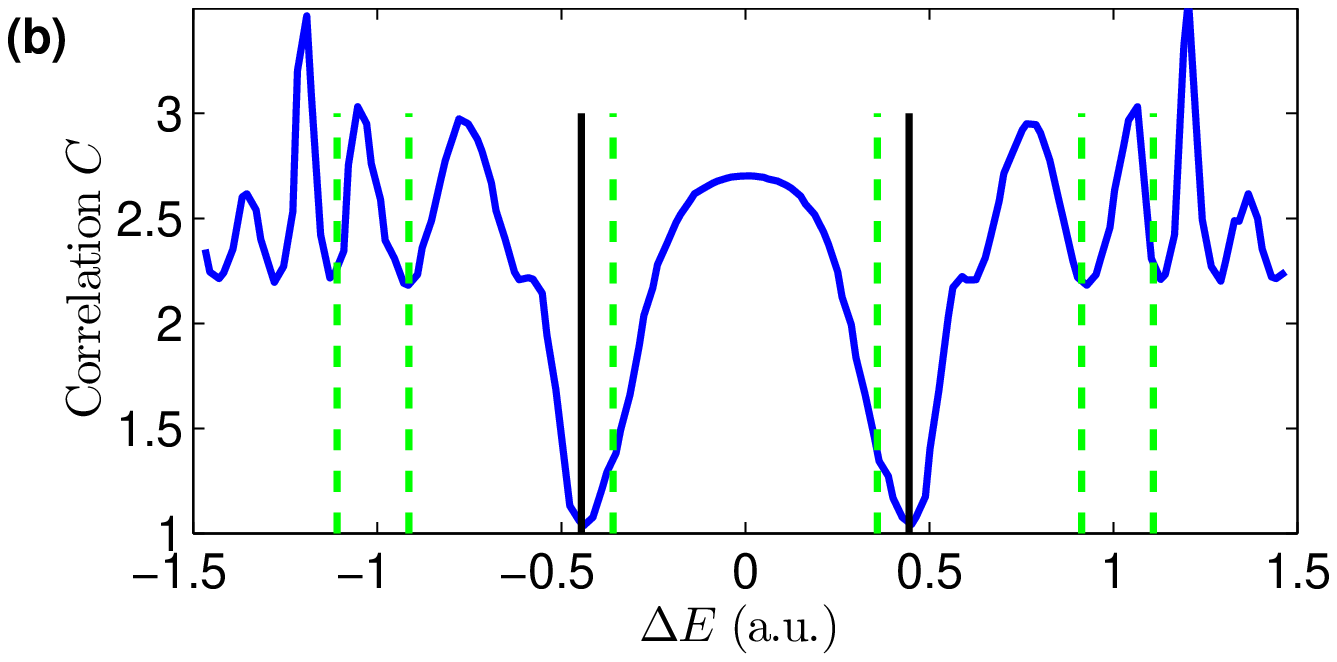}
% from outputCapp/175540
\caption{\label{fig:XUVcorrelation}Top: 
Probability distribution $P(E_{1},E_{2})$ shown in Fig.~\ref{fig:XUVenergyDist}
along the $N=3$ photon line as a function of the energy
difference $\Delta E:=E_{1}-E_{2}$.
Bottom: Angular correlation $C$, Eq.~(\ref{eq:correlationmeasure}), along the same line.
Dashed lines: sequential peaks involving the ionic ground state (solid),
and excited ionic states (dashed).
The minima for higher ionic excitations are slightly 
displaced due to the cutoff of the Coulomb potential at $R_c=20\au$.
}
\end{figure}

In  Fig.~\ref{fig:XUVcorrelation}(a) we show the probability
distribution evaluated along the $N=3$ photon shared energy line of
Fig.~\ref{fig:XUVenergyDist}. 
The JADs in Fig.~\ref{fig:XUVjads} correspond to the points $\Delta E:=E_1-E_2=0$
and $\Delta E\approx 0.45\au$.
Apart from the purely sequential peaks at 
$(E_{1},E_{2})=(-I_{p}^{(1)}+\hbar\omega,-I_{p}^{(2)}+2\hbar\omega)$,
there are several more peaks corresponding to excited states of the  $\mathrm{He}^{+}$ ion.
Denoting by $\mathcal{E}_{n}$ the excitation energy from the ionic ground
state to the $n$-th excited state, the DI efficiency
is enhanced at energies 
$(E_{1},E_{2})=(-I_{p}^{(1)}+2\hbar\omega-\mathcal{E}_{n},-I_{p}^{(2)}+\hbar\omega+\mathcal{E}_{n})$
with $n\in\mathbb{N}$, also see~\cite{Pazourek2011}. As the photon
energy $\hbar\omega\approx1.54\au$ is nearly resonant with the
first excitation energy $\mathcal{E}_{1}=1.5\au$, the resonant and the purely sequential peaks 
are barely discernible.

The degree of correlation along the three photon shared
energy line features a minimum for each maximum of the DI probability.
As expected, correlation is reduced when the transition goes through an intermediate
state that disentangles the detachment of the two electrons.
In particular the value for no correlation $C=1$ is almost reached at the sequential
point $\Delta E=0.45$, where the angular distribution is well
described as  $\propto |Y_{2}^{0}Y_{1}^{0}|^2$, Fig.~\ref{fig:XUVjads}(b).

\section{Double ionization at infrared wavelengths\label{sec:resultsIR}}

%We now present results at the Ti:sapphire wavelength of $780\nm$,
%in particular the ''knee'' of Helium DI.  

%possible additional stuff to do:
%\begin{itemize}
%\item analyze the values where maximum recollision $3.2U_{p}$ crosses the potential energies
%($I_{t1}=3\cdot10^{14}\Wcm$ and $I_{t2}=2.25\cdot10^{14}\frac{W}{cm^{2}}$
%for $780nm$)
%\item verify 5.3 Up cutoff, see~\cite{Liu2008} and~\cite{Parker2006}.
%\item add more direct comparisons with experiments:\\
%\cite{Staudte2007} 800nm 4.5e14W/cm2 (in particular a spectrum in z-dirction)\\
%\cite{Rudenko2007} 800nm 15e14W/cm2
%\end{itemize}

\subsection{Two-color XUV-IR double emission\label{sec:xuv-ir}}

To the best of our knowledge, the only published results of \textit{ab
initio} computations for DI photo-electron \textit{spectra} at 
IR wavelengths treat situations, where an XUV
pulse initiates dynamics and a weak IR pulse
controls the ionization of the excited system. 
For example, in~\cite{Liu2014} an IR pulse with intensity $3\mycdot10^{12}\Wcm$ 
was used to modify the angular distributions of single- and two-photon double
ionization by an XUV pulse.
In~\cite{Parker2003} time-dependent ionization yields
at large intensities were computed, but the used method did not allow for computation of 
DI spectra.

\tempsubsection{Hu2013}

%Progress at higher intensities has only been made recently.
In~\cite{Hu2013} an attosecond XUV pulse with photon energy $\hbar\omega=1.5\au$
tuned to the lowest $\mathrm{He}^{+}$ transition energy was used to
enhance photo absorption from a single cycle IR dressing pulse of moderate
intensity $2\mycdot10^{14}\Wcm$. It was found, that at
time delays between IR field and XUV pulse coinciding with recollision
events, excessive absorption of IR photons is induced by the strong
electron-electron correlation. 

\begin{figure}[H]
\includegraphics[width=1\columnwidth]{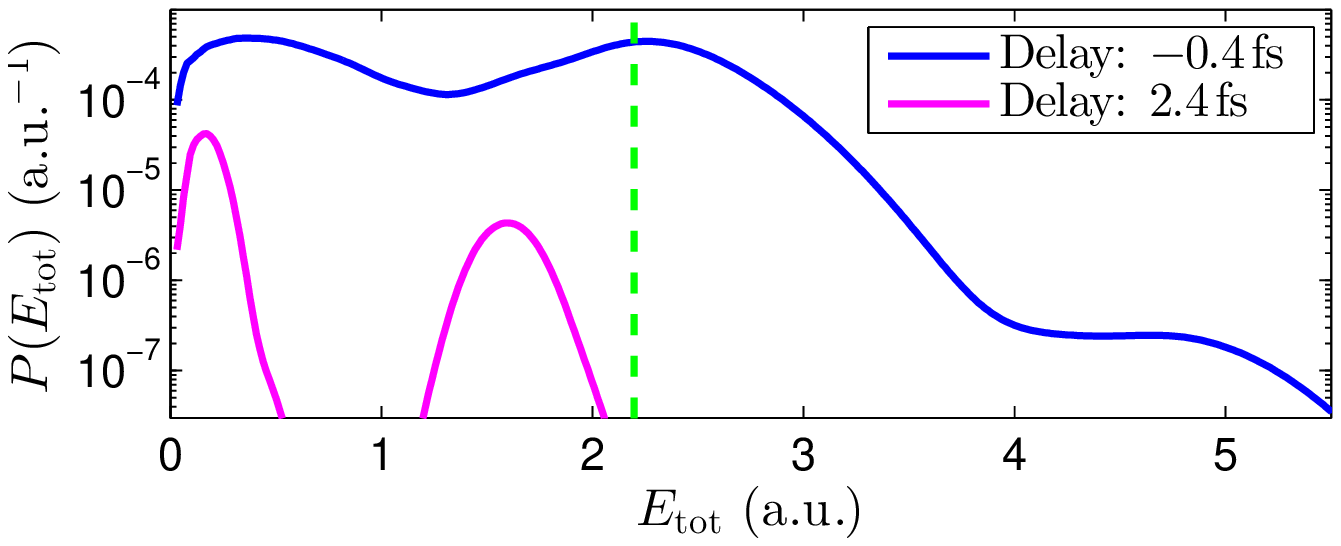}\\
\includegraphics[width=1\columnwidth]{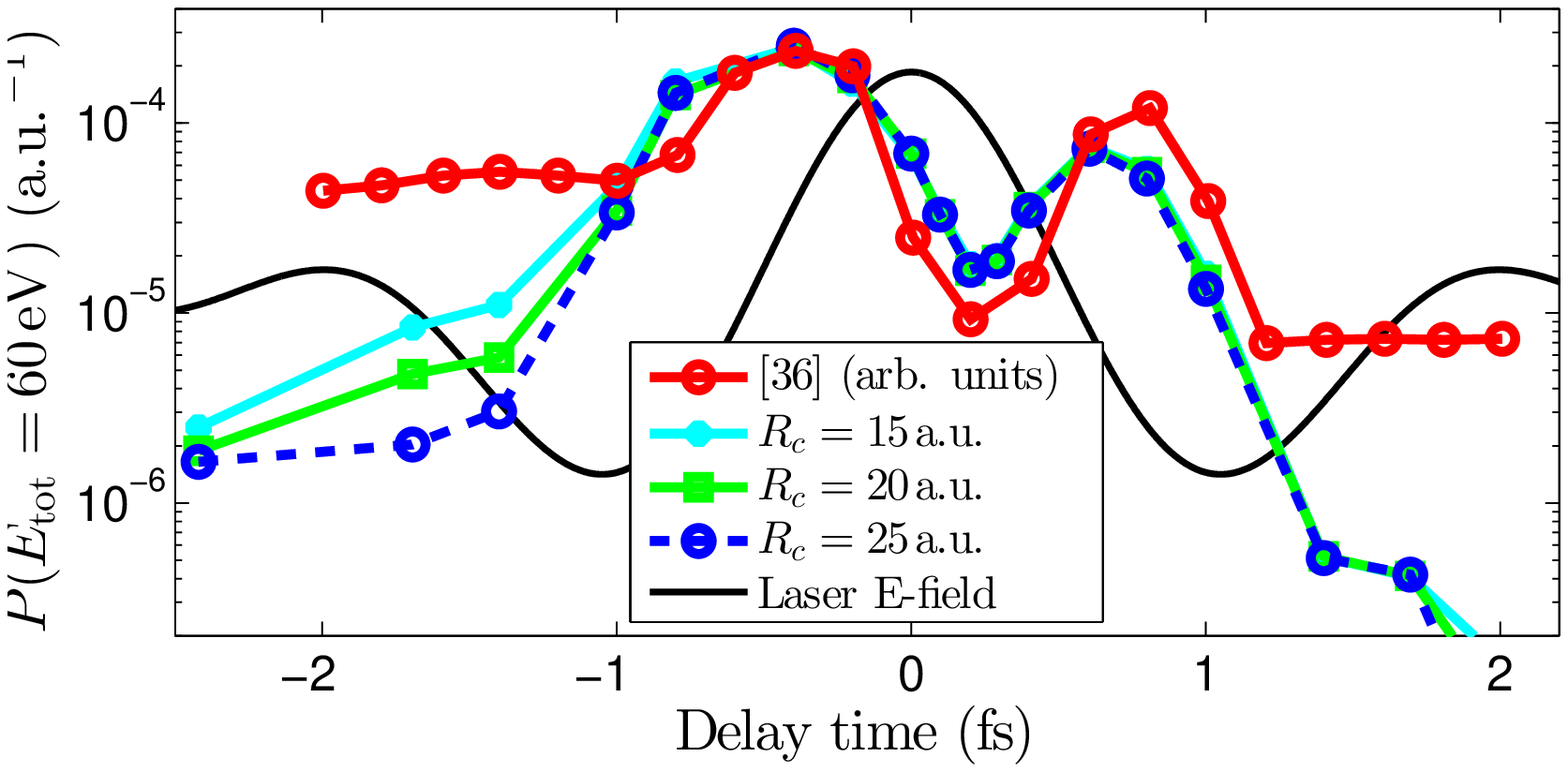}
\caption{\label{fig:hucomp}Top: 
Photo-electron spectrum as a function of the shared total energy,
Eq.~(\ref{eq:PEtot}),
for two relative time delays between XUV and IR pulses.
Large positive time delays correspond to the XUV pulse coming after the IR pulse.
Green dashed line indicates $E_\mathrm{tot}=60\eV$.
Bottom: $P(E_\mathrm{tot}=60\eV)$ as a function of the relative time delay: dependence on $R_c$
and comparison with Ref.~\cite{Hu2013}. Arbitrary units 
in~\cite{Hu2013} are adjusted to 
approximately match our results. 
}
\end{figure}

Fig.~\ref{fig:hucomp} shows our results for the probability density 
at total emission energy $E_{\rm tot}=E_1+E_2$
\begin{equation}
P(E_\mathrm{tot})=
\int_0^{E_\mathrm{tot}}\mathrm{d}E_1\,P(E_1,E_\mathrm{tot}-E_1).
\label{eq:PEtot}
\end{equation}
The upper pane shows the enhancement across the whole energy range.
The lower pane singles out $E_{\rm tot}=60\eV$ as a function of delay time.
We reproduce the overall picture reported in Ref.~\cite{Hu2013}, but find significant
quantitative discrepancies. 
Note that the comparison is not in absolute numbers,
as results of~\cite{Hu2013} are given in arbitrary units.
For example, at larger positive offsets Ref.~\cite{Hu2013} shows 
nearly constant data points for large positive offsets, which does 
not match with our computations. Such a behavior may appear implausible,
as at these delays the IR has nearly passed when the XUV arrives, and yields 
should fall to the very low level of pure XUV double ionization at $E_{1}+E_{2}=60\eV$.
We would like to remark that the simulation box size of $305\au$ used 
in~\cite{Hu2013} falls short of the distance of $\gtrsim 400\au$ that 
$60\eV$ electrons travel during the IR pulse duration. For tSurff, 
the cutoffs at $R_c\leq 25\au$ used in our simulations would mask
long range Coulomb and post-collision effects, however the impact of $R_c$ 
appears to be small, see Fig.~\ref{fig:hucomp}.

%168508 hu780compare.m

\hide{
\commAS{mabe move: 
In these calculations, the comparatively low resource requirement of tSurff is clearly 
exposed. Each of our computations included a total of $551$ partial waves 
and was conducted on a 16 core machine, running
for a maximum of two days. Compared to the over 4000 cores
of a Supercomputer utilized for each run including $295$ partial waves in~\cite{Hu2013},
this is a dramatic reduction of spent computational resources.
The simple reason for this gain is the small box size of only $R_c=25\au$ 
compared to $305\au$ in Ref.~\cite{Hu2013}. 
As can be seen in Fig.~\ref{fig:hucomp}(b), with tSurff even 
box sizes as small as $15\,a.u.$ qualitatively satisfactory results are obtained.
}
}
\subsection{Double ionization enhancement by recollision\label{sec:DIrecollision}}

% The full fragmentation of Helium by an IR laser pulse \textit{alone} is only fully understood
% for very high laser intensities $\gtrsim 5\mycdot10^{15}\Wcm$,
% where the description by two independent and sequential
% tunnel ionization events agrees with experiment~\cite{Walker1994}.
% At lower intensities, predictions by such a sequential model are wrong by several orders of magnitude.
% It has been widely accepted that in this case field induced recollisions
% lead to some form of non-sequential double ionization (NSDI)~\cite{Ishikawa2010}.
% Several such NSDI pathways have been proposed,
% including simultaneous ejection (SE)
% and
% recollision induced excitation with subsequent ionization (RESI)~\cite{Haan2006,Shaaran2010}.
% For kinetic energies of the recolliding
% electron below the ionic excitation threshold $\mathcal{E}_1=1.5\au$,
% which is the case for intensities below $\approx 2.3\mycdot10^{14}\Wcm$ at $780\nm$,
% these pathways are closed and a mechanism called doubly delayed ejection (DDE)~\cite{Emmanouilidou2011}
% seems to dominate.
% %For a quantitative inspection the necessary accurate and detailed benchmark data is still lacking.
% %Full understanding of the DI processes is still lacking,
% %it is even unclear if there are additional, until now unthought of mechanisms. 

To this date, double ionization at IR wavelength has not been reproduced 
by solutions of the two-electron TDSE for the the full intensity range because of 
the high demand on computational resources, 
see section~\ref{sec:ProblemHelium}.
Using tSurff, we can provide yields up to intensities  $4\times10^{14}\Wcm$,
with error estimates of $\lesssim  20\%$ up  $3.5\times10^{14}\Wcm$,
using only moderate computational resources.
 
\subsubsection{Double to single ionization ratio}

\begin{figure}[h]
\includegraphics[width=1\columnwidth]{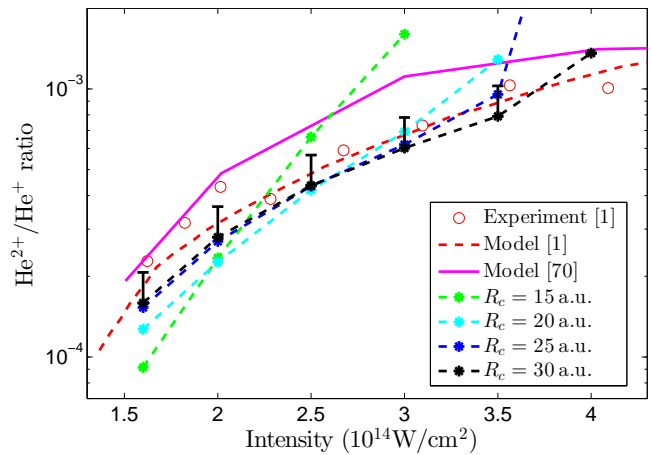}
\caption{\label{fig:Ratio-of-double}
Ratio of DI yield and single
ionization yield as a function of intensity. Experiment and model \cite{Walker1994}, 
model calculations \cite{Ishikawa2010}, and our full 2-electron results for increasing $R_c$.
Upward error bars indicate the long pulse limit.
}
\end{figure}

The ratio of double to single ionization yields is shown in Fig.~\ref{fig:Ratio-of-double} 
for $R_c$ up to $30\au$ and compared to experiment. 
Except for $R_c$, our results are converged with respect to all other discretization 
parameters to within a few percent. We used a laser pulse with single
cycle rampup, $n$ cycles at full intensity and one cycle rampdown at wavelength $\lambda=c\tau=780\nm$:
\begin{equation}
A(t)\propto f_{\al,\be}(-t)\sin\big(2\pi t/\tau\big)f_{\al,\be}(t).
\label{eq:laserPulseShape780-1}
\end{equation}
with $\al=n\tau/2$, $\be=(n/2+1)\tau$ and $f_{\al,\be}$ as in 
Eq.~(\ref{eq:truncationinterval}).
The calculations were performed with pulse durations of $n=4$ cycles. 

Fig.~\ref{fig:Ratio-of-double} shows the double-to-single ratio as obtained
with $R_{c} = 15,20,25$ and $30\au$ and smoothing $R_c-\al=3,4,6$ and $8\au$, respectively, 
see Eq.~(\ref{eq:truncationinterval}). One can clearly see larger intensities require 
larger $R_c$, which roughly correlates with the quiver radius $Q$. In the intensity range
$1.6-3\times10^{14}\Wcm$ with quiver radii $Q=20-27\au$ results vary by at most $25\%$ between $R_c=20$ and $R_c=30$. 
At the intensities $I\geq3.5\times10^{14}\Wcm$, numerical results for $R_c<30$ strongly depart 
from our largest calculation with $R_c=30$. 
The lower intensity limit for our calculations is  $\lesssim 1.5\mycdot 10^{14}\Wcm$, where
the overall yields are so small that numerical inaccuracies render the results useless.

There is some dependence of our results on the pulse duration: 
using a two-cycle ramp up we found changes of less than $5\%$ at 
selected intensities. More important is the dependence on pulse duration.
At the intensities $\leq 2.5\times10^{14}\Wcm$ we found an increase of 
ratio for $n=4,5,\ldots$ cycles which saturates at about 30\% for $n\approx8$ with no
relevant further increase for durations up $n=12$. We expect similar pulse duration 
dependence at higher intensities. This increase by 30\% is indicated by an (upward) error bar
for our $R_c=30$ calculation in Fig.~\ref{fig:Ratio-of-double}.

Fig.~\ref{fig:Ratio-of-double} also shows the experimental results and model predictions.
There is good agreement with the experimental data, but that may well by fortuitous, 
as experimental intensities were subjected to errors as large as $30\%$.
The analytical model of Ref.~\cite{Walker1994} is based on the single active electron approximation (SAE)
and on the ac-tunneling (ADK) rates~\cite{Ammosov1986} and is, by its construction, close to the
experimental data. 
The model in~\cite{Ishikawa2010} implements the rescattering scenario using the 
SAE approximation in combination with electron-ion impact cross sections.
It appears to somewhat overestimate the actual ratio, as
it remains outside our estimated tSurff error.
Note that in the plot shown in  Fig.~1 of Ref.~\cite{Ishikawa2010}, experimental intensities
are scaled by a factor 1.15 for the purpose of the comparison.

\subsubsection{Energy probability distributions}

\begin{figure}[H]
\includegraphics[width=0.5\columnwidth]{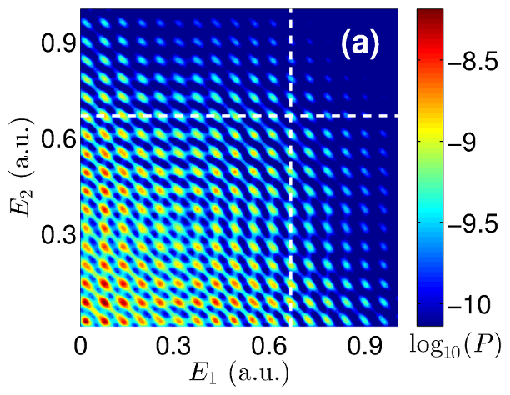}\includegraphics[width=0.5\columnwidth]{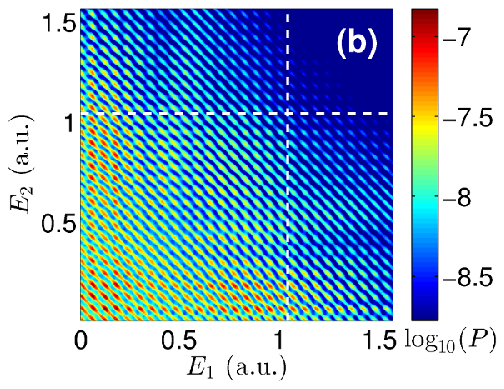}\\
\includegraphics[width=0.5\columnwidth]{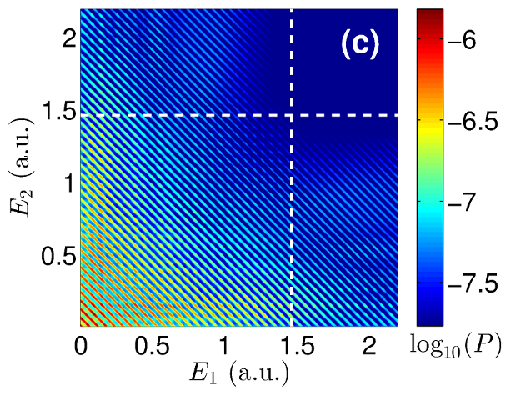}\includegraphics[width=0.5\columnwidth]{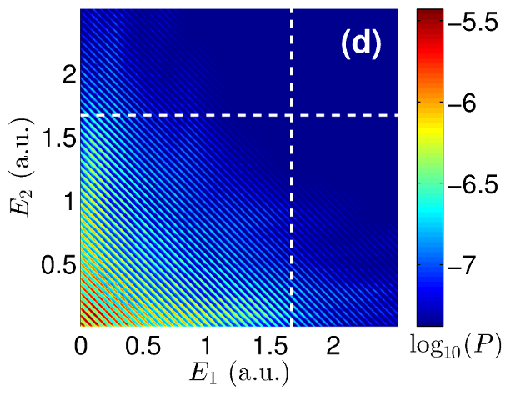} 
\caption{\label{fig:780DIspecs}
Total DI spectra $P(E_{1},E_{2})$, Eq.~(\ref{eq:energyprobdist}),
for several of the data points with truncation interval $R_c=30\au$ shown in Fig.~\ref{fig:Ratio-of-double}.
Intensities in units of $10^{14}\Wcm$ are  (a) $1.6$, (b) $2.5$,
 (c) $3.5$ and (d) $4.0$. White dashed lines mark $2\,U_p$.
}
\end{figure}

In Fig.~\ref{fig:780DIspecs} we present energy probability distributions $P(E_1,E_2)$ 
for IR double ionization at intensities $I=1.6, 2.5, 3.5$ and $4\,\Wcm$, 
with the pulse shape as in Fig.~\ref{fig:Ratio-of-double} for photo-electron energies $E_1,E_2<3\,U_p$.
One sees that the 4-cycle pulses define the carrier frequency $\omega$ well enough
to clearly distinguish individual DI-ATI peaks separated by the photon energy $\hbar\omega$.
One also observes changes in the DI emission pattern with increasing intensity $I$.
In Fig.~\ref{fig:780DIspecs}(a) and (b) conspicuous enhancement of double emission in the
area $E_1\approx E_2\approx 1.7\, U_p$ appears. Note that at the corresponding intensities $1.6$ and $2.5\mycdot10^{14}\Wcm$
the maximal recollision energy remains below the second ionization potential of $2\au$.
As recollision cannot be the sole DI mechanism, one may speculate that processes like 
simultaneous tunneling and doubly delayed emission (DDE)~\cite{Emmanouilidou2011} where 
final energies $E_1$ and $E_2$ are comparable play a greater relative role. 
At (c) and (d) direct excitation by the recolliding electron becomes accessible and a roughly 
L-shape energy distribution emerges.

%relativeCap = 1e-2; 158275 , 158605 , 163073 , 
% we may exchange the SMALL_... version by the version without SMALL_.
% this requires adaption of tarit file
% SMALL_ files are created by opening original .eps file with gimp, then import with 300dpi, then "image->scale image" to 2 inches, then save.

For estimating the $R_c$-induced errors of the energy distributions
we define the relative difference $\mathcal{E}$ between two distribution $P_a$ and $P_b$ as
\begin{equation}
\mathcal{E}(E_{1},E_{2}):=\frac{|P_{a}(E_{1},E_{2})-P_{b}(E_{1},E_{2})|}
{\overline{P}_{\hbar\om}(E_{1},E_{2})},
\end{equation}
where $\overline{P}_{\hbar\om}(E_{1},E_{2}):=\max_{E_1',E_2'} P(E'_{1},E'_{2})$ denotes
the maximum over a $\hbar\om$-neighborhood
with $(E'_1-E_1)^2+(E'_2-E_2)^2<(\hbar\om)^2$.

\begin{figure}[H]
\includegraphics[width=0.49\columnwidth]{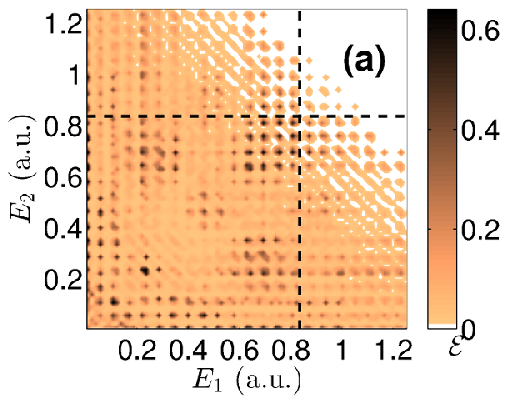}
\includegraphics[width=0.49\columnwidth]{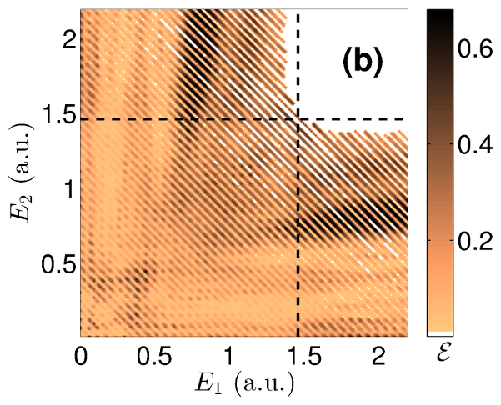}
\caption{\label{fig:relErrSpec}
Error estimates $\mathcal{E}$ by comparing computations with (a) 
radii $R_c=25\au$ vs.\ $R_c= 35\au$ for $I=2\mycdot10^{14}\Wcm$,
and (b) $R_c=25\au$ vs.\ $R_c= 30\au$ for $I=3.5\mycdot10^{14}\Wcm$.
Errors are only plotted where $P(E_1,E_2)$ is larger than $1\%$ of its maximum.
}
\end{figure}

Two examples are shown in Fig.~\ref{fig:relErrSpec}.
At $I=2\mycdot10^{14}\Wcm$ with quiver radius $Q=22\au$ 
 we compare calculations with $R_c=25\au$ and $R_c= 35\au$, Fig.~\ref{fig:relErrSpec}(a).
Relative differences approach $60\%$ where one electron has low energy and also near the energy
diagonal. These regions can be expected to be strongly affected by Coulomb truncation.
In many areas differences remain below $20\%$.
The intensity $I=3.5\mycdot10^{14}\Wcm$ is near the limit of our presently accessible parameter 
range, Fig.~\ref{fig:relErrSpec}(b). At this intensity 
the quiver radius $Q=29\au$ exceeds
the truncation radius $R_c=25$. Still, comparing to $R_c= 30\au$, relative differences 
approach $60\%$ only at few places in the relevant region $E_1+E_2\leq 2\,U_p$, 
and are mostly below $30\%$. Large relative differences  
naturally appear where yields become small. 

\begin{figure}[H]
\includegraphics[width=0.49\columnwidth]{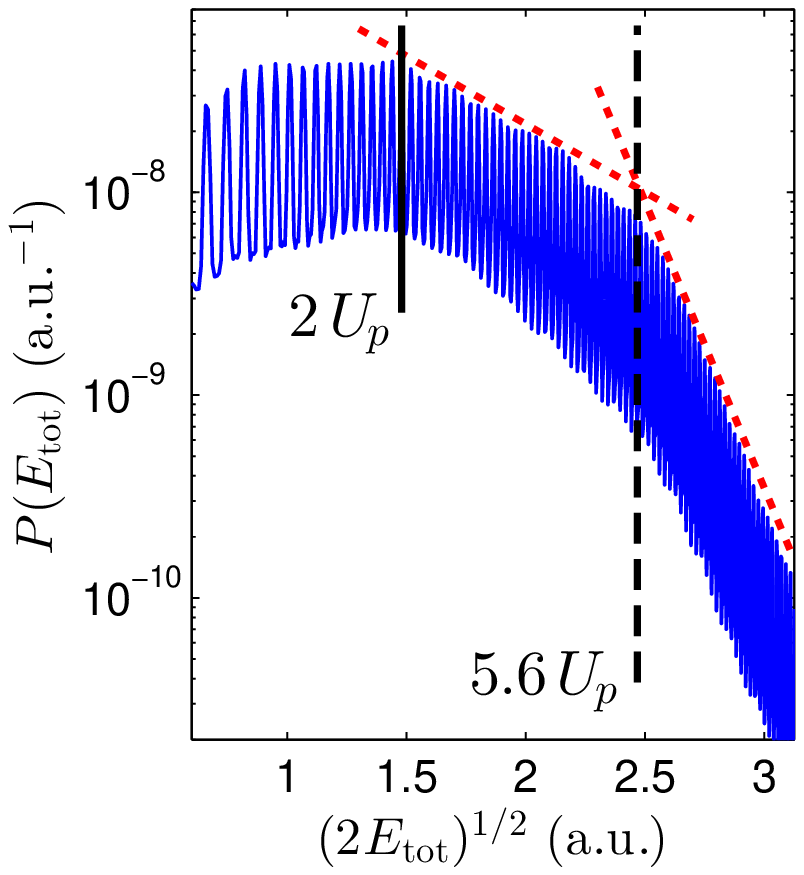}
\includegraphics[width=0.49\columnwidth]{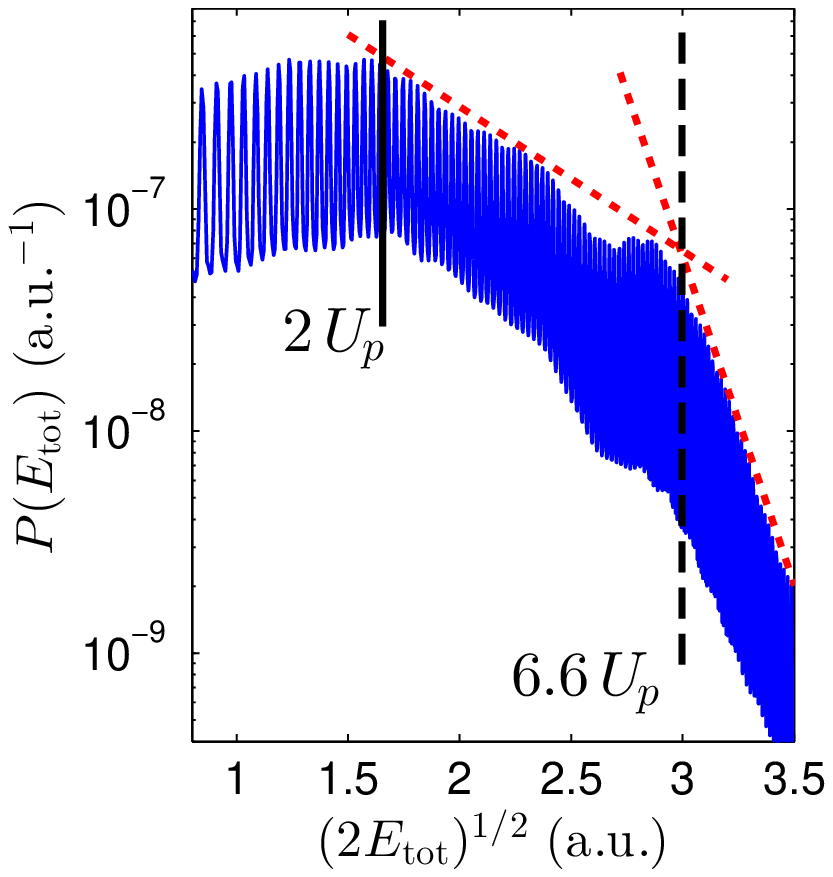}
\caption{\label{fig:parkerCompare}
Total momentum spectrum $P(E_{\rm tot})$, Eq.~(\ref{eq:PEtot}),
for a $780\nm$ $n=4$ cycle pulse.
Left: $2.6\mycdot 10^{14}\Wcm$ cutoff at $5.6\,U_p$.
Right: $3.25\mycdot 10^{14}\Wcm$ cutoff at $6.6\,U_p$.
Solid lines mark the momentum corresponding to $2\,U_p$, dashed lines mark the transition to the exponential 
decay at the cutoff of the spectrum, dotted lines indicate the slopes. 
The behavior conforms with Fig.~3 of~\cite{Parker2006} for
$390\nm$ at the corresponding intensities $1.04\mycdot 10^{15}\Wcm$
and $1.3\mycdot 10^{15}\Wcm$.
}
\end{figure}

In~\cite{Parker2006}, a sharp transition of the cutoff in the shared DI energy distribution 
from $5.3\,U_p$ to $>\!\!7\,U_p$ was found, when maximum recollision energies surpass 
$I_{p}^{(2)}=2\au$ at $\lambda = 390\nm$. 
A similar cutoff was reported in~\cite{Liao2010} for a one-dimensional model. 
We could reproduce this for $390\nm$ (not shown) and find the cutoff also in full three dimensions
at $780\nm$, see Fig.~\ref{fig:parkerCompare}. 

%381575 and 384361

\subsection{Angular distribution}

\begin{figure}[H]
\begin{center}
\includegraphics[width=0.48\columnwidth]{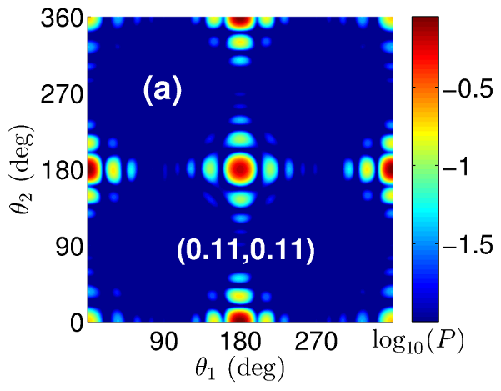}
\includegraphics[width=0.48\columnwidth]{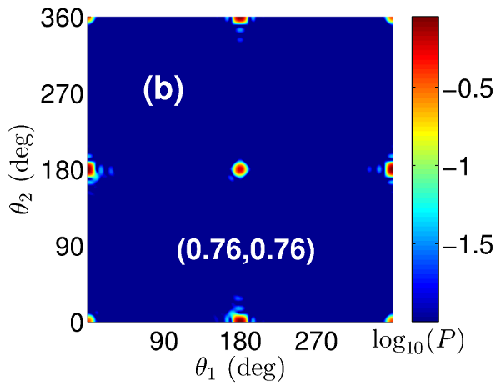}\\
\includegraphics[width=0.48\columnwidth]{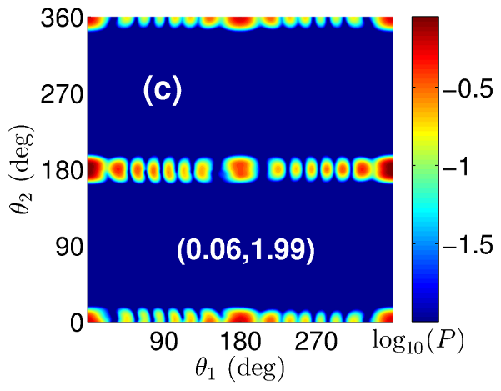}
\includegraphics[width=0.48\columnwidth]{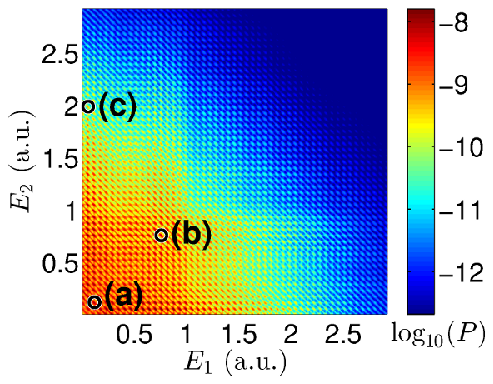}
\end{center}
\caption{\label{fig:780_2e14_overview}
JADs for wavelength $780\nm$ and intensity $I=2\mycdot10^{14}\Wcm$.
The panes (a), (b), and (c) correspond to the energies marked in 
the $P(E_{1},E_{2})$ distribution, lower right pane. Plots are
normalized to $\max_{E_1,E_2}[P(E_1,E_2)]=1$.
}
\end{figure}

JADs depend most sensitively 
on $R_c$ and full convergence could not be achieved with moderate 
computational effort. 
In Fig.~\ref{fig:780_2e14_overview} we show JADs for three different $(E_1,E_2)$ at
$I=2\mycdot10^{14}\Wcm$.
In (b) it can be seen that if both electrons escape with large energies the angular emission
pattern is uncorrelated and highly focused around the polarization axis.
If one electron barely manages to escape, then its angular
distribution is less focused and the JADs exhibit complex structures, Fig.~\ref{fig:780_2e14_overview}(c).
Correspondingly, if both electrons leave with small energies we observe correlated angular
emission patterns, an example of which can be seen in (a).

%192108. cap 1e-2. xlim({[}0.01 1.1{]}).

\begin{figure}[H]
\includegraphics[width=1\columnwidth]{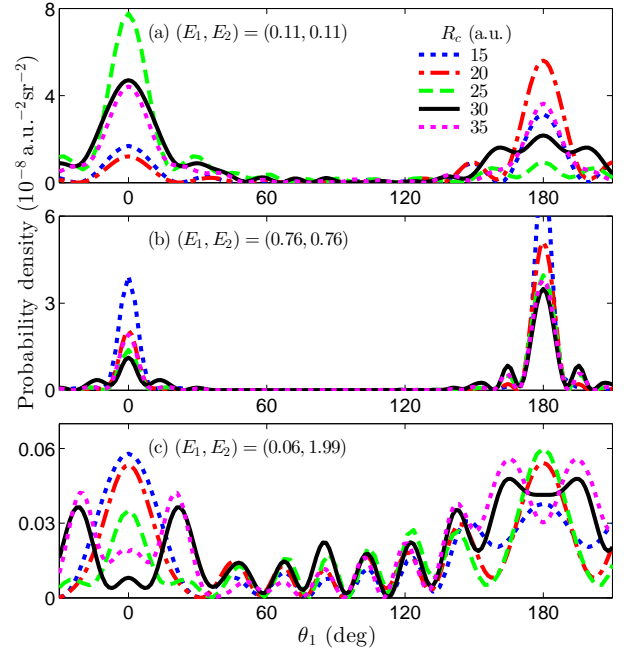}
\caption{\label{fig:jadConvergenceWithRc}
Cuts through JADs of Fig.~\ref{fig:780_2e14_overview} at $\theta_2=0$ for 
various values of $R_{c}$.
At low energies convergence is only qualitative, see (a) and (c).
At high energies the angular distribution stabilizes at $R_c = 30\au$, 
see (b). 
}
\end{figure}

The JADs shown in Fig.~\ref{fig:780_2e14_overview} are accurate with respect to the angular
momentum expansion to the level of $\sim\! 10\%$.
Convergence with $R_c$ depends on the final electron energies.
In Fig.~\ref{fig:jadConvergenceWithRc} we present cuts through JADs of Fig.~\ref{fig:780_2e14_overview}
at $\theta_2=0$.
If both energies are large, then neither the exact nuclear potential shape nor
postcollision interactions are relevant, and convergence is achieved with $R_c=30\au$,
Fig.~\ref{fig:jadConvergenceWithRc}(b).
However, if at least one of the electrons energies is small,
then electrons may interact with the nucleus over long times
and convergence with $R_c$ could only be achieved to the level of qualitative
agreement. The overall distribution does not change completely but some
qualitative features are still in flux. For example at the low emission energies $E_1=E_2=0.11$, 
Fig.~\ref{fig:jadConvergenceWithRc}(a), the dominant emission direction
changes from back-to-back to side-by-side when increasing $R_c=20\au$ to
$R_c = 25\au$. In (c) local minima appear along the polarization axis 
with $R_c = 30\au$.

Clearly, larger $R_c$ are required for convergence at low energies.
The present implementation of the method renders such calculations impractical with
reasonable computational resources. Also, while for $2\mycdot10^{14}\Wcm$ computations with $R_c=35\au$
were still accomplishable, at higher intensities the increasing demand on angular momenta is prohibitive
and only computations with up to $R_c=30\au$ were practical.

\subsection{Angular correlations}

As for XUV double ionization  we compute the 
angular correlation by Eq.~(\ref{eq:correlationmeasure}),
and find it to follow the DI-ATI structure, although not as clearly as in Fig.~\ref{fig:XUVcorrelation}).
In addition, at IR wavelength we find maxima at small electron escape energies $E_{1,2}\lesssim 0.2\au$.
Strong correlation at low energies is to be expected as the electrons interact over longer times before leaving
the vicinity of the nucleus.

\begin{figure}[H]
\includegraphics[width=0.5\columnwidth]{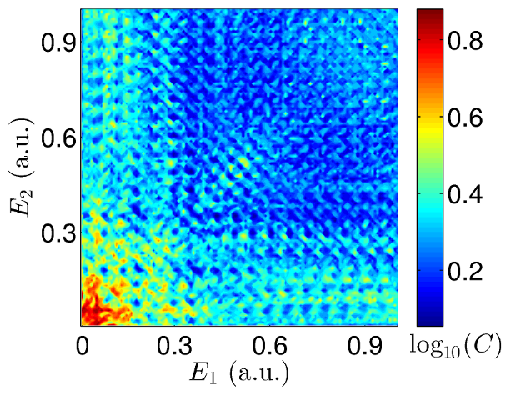}\includegraphics[width=0.5\columnwidth]{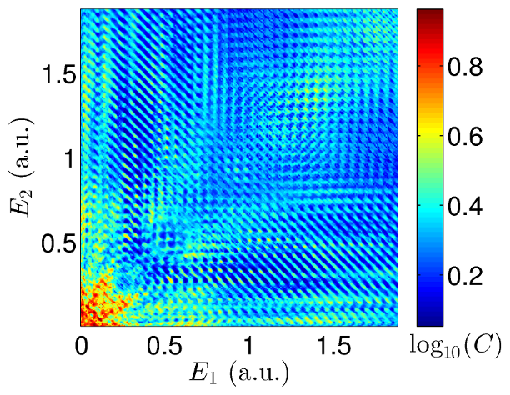}
\caption{\label{fig:IRcorrelation}
Angular correlation $C$ as a function of photo-electron energies for intensities 
$1.6\mycdot10^{14}\Wcm$ (left) and $2.0\mycdot10^{14}\Wcm$ (right).
Computations with $R_c=30$.}
\end{figure}

\subsection{Correlated momentum distribution}
Measurements of differential double-emission spectra at IR wavelength 
were reported in Ref.~\cite{Rudenko2007} 
for $I>10^{15}\Wcm$ and at the somewhat lower intensity of  $I=4.5\mycdot10^{14}\Wcm$
in Ref.~\cite{Staudte2007}. 
In~\cite{Staudte2007} a ''fingerlike'' structure was found 
in the photo-electron momentum distribution for side-by-side momenta 
on the polarization axis $\th_1=\th_2=0$ and $\th_1=\th_2=\pi$.
The structure is manifest at energies $>2U_p$, where it shows a minimum at $E_1=E_2$ 
and emission maxima off the energy diagonal. 
It is attributed to
electron-impact ionization with backscattering
at the nucleus upon recollision, analogous to the recoil peak appearing
in field-free ionization by a scattering electron~\cite{Chen2010a}. Similar structures
were seen in a 1+1-dimensional model calculation \cite{Liao2010}.

\begin{figure}[H]
\centering
\includegraphics[width=0.7\columnwidth]{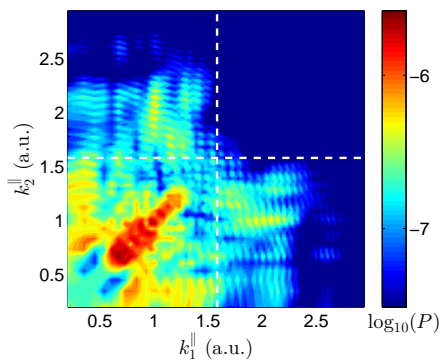}
\caption{\label{fig:spectrumAlongZ} %166229
Correlated side-by-side emission for DI at $780\nm$, $I=3\mycdot10^{14}\Wcm$,
for momenta parallel to the polarization axis.
Computation with  $R_c=30\au$. Dashed lines mark $2\,U_p$,
beyond which ``fingerlike'' structures appear 
similar as observed in experiment~\cite{Staudte2007}.
}
\end{figure}

The laser intensities used in \cite{Staudte2007}
are just beyond the limitations of the present implementation of tSurff.
At wavelength $780\nm$ and the somewhat lower intensity of $I=3\mycdot10^{14}\Wcm$ 
recolliding electrons can still directly ionize
the parent ion, the DI mechanism proposed in \cite{Chen2010a} still functions,
and one expects similar structures as reported in  \cite{Staudte2007}.
Fig.~\ref{fig:spectrumAlongZ} shows that this is indeed the case.
In particular, we reproduce the minimum along the diagonal of equal momenta.
Due to the sensitivity of angle resolved observables to $R_c$, especially
for side-by-side emission, we expect details of our result to change for  $R_c>30$.

\subsection{Convergence with spatial discretization\label{sec:convergenceOfDisc}}

The most important spatial convergence parameter is the number of partial waves in 
Eq.~(\ref{eq:discretization}).
We illustrate this for fixed tSurff radius $R_c=20$. 

With a linearly polarized pulses, the  $M=0$  magnetic quantum number
of the He ground state is conserved such that $m_1=-m_2=:m$.
The population of the $m$-components of $\Psi$ changes only by electron collisions
and we found rapid convergence to within our desired precisions at values $m\lesssim4$.

In contrast, $l_1$ and $l_2$ are directly populated by the laser interaction. 
In the perturbative regime angular momenta remain small overall,
and no particular constraints
beyond a simple square $l_{1},l_{2}\leq l_{\rm max}$ were applied. Also, as overall problem sizes
remain comparatively small in the perturbative regime, we made no attempt to rearrange the partial waves
into eigenspaces of total angular momentum.

\begin{figure}[H]
\centering
\includegraphics[width=0.7\columnwidth]{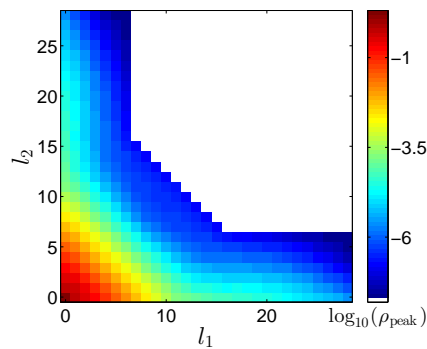}
%partialwavemasses.m 163763
\caption{\label{fig:angularmomentumgrid} 
Peak partial wave populations 
$\rho_{\rm peak}(l_1,l_2)$ for a pulse with $\lambda=780nm$, $n=4$ and $I=2\mycdot10^{14}\Wcm$.
}
\end{figure}

In the non-perturbative regime partial waves reach large $l_1,l_2$, however 
mostly in an L-shaped area of the $l_{1}$-$l_{2}$-plane.
In Fig.~\ref{fig:angularmomentumgrid} we show 
the peak partial wave populations of the $m=0$ component of $\Psi(t)$
\begin{equation}
\rho_{\rm peak}(l_{1},l_{2}):=\max_{t}||\l Y_{l_1}^0 Y_{l_2}^0|\Psi(t)\r||^{2},
\end{equation}
as reached at any time during the interaction with a 4-cycle laser pulse 
with $\la=780\nm$ and $I=2\times10^{14}\Wcm$.
Similar patterns are found for larger $m$. 
Based on this observation, we selected the $(l_1,l_2)$ from an L-shaped region adjusted by inspecting 
the populations $\rho_{\rm peak}(l_1,l_2)$ at the borders.
Comparing calculations with  551, 639, and 737 partial waves,
we find the JADs converged to $< 10\%$ almost everywhere. We observed larger differences $\lesssim 30\%$
only near back-to-back and side-by-side emission. 

\tempsubsection{//finite elements}

Convergence of the radial expansion is achieved quite easily:
for $R_c=20\au$ typically four radial sections $r_i\up{n_i}-r_i\up{n_i-1}=5\au$ 
were used with degree 
11 polynomials at the first element, degree $7$ 
up to  $r_i\up{4}=20\au$ and 
exponentially damped polynomials of maximal degree 19 for absorption in the last section, c.f.\ Eq.~(\ref{eq:radialProductR}).
The higher degree in the first element accelerates convergence of the two-electron 
bound state. Such a discretization gives JADs which are radially converged 
within the accuracies discussed here.

It may be possible to further reduce the radial basis by making the discretization of each
$[r_1\up{n_1-1},r_1\up{n_1}]\times[r_2\up{n_2-1},r_2\up{n_2}]$ patch dependent on its position.
%in a similar as used for $(l_1,l_2)$.
For example at large $r_1$ and small $r_2$, the solution in $r_2$-direction will resemble 
the lowest ionic bound states which can be parametrized by very few functions. Such constraints
were not explored here, but may be useful for further pushing the limits of the calculations.

\section{Summary and Outlook}

We have presented calculations of laser double ionization of 
the Helium atom in a wide range of intensities and wavelengths. Both, XUV and 
near IR wavelengths were covered. For the majority of our results we can 
provide well-founded accuracy estimates. 

Our calculations reproduce key literature results at XUV wavelength.
As a challenging example we have chosen XUV two photon double ionization,
where we reproduce available literature results.
In the most disputed region near the 
single-photon ionization threshold we agree with Refs.~\cite{Feist2008,Palacios2010,Nepstad2010}.

At the IR wavelength of $780\nm$ we computed {\it ab initio}
the double-to-single ionization ratio of He. 
Results conform with experiment and with a numerical recollision model~\cite{Ishikawa2010},
although convergence indicates that Ref.~\cite{Ishikawa2010} may somewhat
overestimate the ratio. We were also able to present differential double ionization
spectra at this wavelength. There are few theoretical results available in literature for comparison.
We qualitatively confirm a two-color XUV-IR result~\cite{Hu2013} but find notable quantitative
deviations. We could extend an observation about a pronounced cutoff in the shared
energy momentum spectra reported for $\la=390\nm$ in Ref.~\cite{Parker2006} to $780\nm$.

Experiments with He at IR wavelength are mostly performed at higher intensities for 
reaching sufficient count rates, where we could not reach convergence in the 
differential spectra with the present implementation. At an intensity slightly below the experimental one,
we qualitatively reproduce the correlations observed in
doubly differential momentum spectra~\cite{Staudte2007}.

All our results are derived from fully differential spectra. Naturally, integrated
quantities converge more quickly. In the XUV regime, satisfactory convergence could 
be obtained for all observables considered. At IR, the double-to-single ratios are accurate to 
within $20\%$ for the 4-cycle pulses used in the simulations up to intensities $3.5\times10^{14}\Wcm$.
Energy differential spectra have typical relative errors of $\lesssim 30\%$, 
which are exceeded only in limited regions of the  $E_1E_2$-plane.
Angle-resolved quantities
are most sensitive to our discretization and for IR the corresponding spectra should only be considered
as qualitative results when post-collision interactions come into play. 

Importantly, calculations presented here are on only moderate computational scale by 
present day standards. At XUV wavelength, a maximum of 16 cores on a single computer node were used.
The largest calculation at IR wavelength
was performed on 128 cores distributed over 8 compute nodes. This should be contrasted
with the use of very large scale computer facilities employed for 
Refs.~\cite{Parker2006,Feist2008,Feist2009,Pazourek2011,Palacios2010,Hu2013}.

This advancement of possibilities is brought about by the tSurff method,
whose potential and limitations we laid out in some detail here. Its main advantage --- the 
numerical simulation on only a small spatial domain --- also entails its main 
shortcoming in the present implementation, namely the inability to reproduce post-collision
interactions that occur outside the simulation domain. Fortunately, for 
a great many of observables and final momenta, post-collision interaction is
of secondary importance. 

There is a range of possible improvements of the approach. On the one hand this concerns 
algorithms and discretization methods. As we could recently demonstrate that FE-DVR methods
are applicable for tSurff~\cite{weinmueller2015}, replacing the present finite element discretization
with FE-DVR may allow significant reduction of the floating operations count
and improve scalability to massive parallel computations. Another algorithmic improvement is
to exploit the low multipole order of the electron repulsion at larger distances 
and the interdependence of radial and angular discretization. 
On the other hand, as has been indicated in section~\ref{sec:tsurffDI}, one may extend 
tSurff theory by using analytic models for reducing the problems due to the truncation 
of the potentials at $R_c$.

By implementing even only part of these measures, full convergence
of the differential information at the present parameters and access to more 
demanding situations such as elliptically polarized IR  pulses, double ionization 
of multi-electron systems, see~\cite{Majety2015a}, or breakup of non-Coulombic systems
and systems with reduced symmetry appear realistic.

We conclude with pointing out that an implementation of tSurff method  in its scalar single electron version
has been made available as a public domain code tRecX~\cite{tRecXcode}.
Publication of the MPI parallel and two-electron features in the same framework is planned for the near future.

\section*{Acknowledgements}
The authors acknowledge financial support by the excellence cluster "Munich Center for Advanced Photonics (MAP)"
and by the Austrian Science Foundation project ViCoM (F41). We are grateful to Stefan Nagele for exentensive
support in detailed comparisons with the TU Wien group's data.

\appendix

\section{Matrix elements $\langle\chi_{\vec{k}}|[H_{V},\Theta]|\psi\rangle$\label{sec:appen:commutator}}
For the commutator matrix element  
appearing in equations~(\ref{eq:Fiswahat}), (\ref{eq:sourceterm}) 
and~(\ref{eq:tsurffInt}) we
express $[H_{V},\Theta]$ in polar coordinates for a $z$-polarized pulse
\begin{align}
\Big[-\frac{\Delta}{2}+\mathrm{i}\vec{A}&\cdot\vec{\nabla},\Theta\Big] = -\frac{1}{2}\delta(r-R_{c})\partial_{r}
\nonumber\\
&-\left(\frac{1}{2}\frac{1}{r^{2}}\partial_{r}r^{2}+\mathrm{i}A_{z}\cos\theta\right)\delta(r\!\!-\!\!R_{c}).
\end{align}
Expanding the plane wave into spherical Bessel functions $j_l$, the Volkov wave becomes
\begin{equation}
\chi_{\vec{k}}(\vec{r},t)=\frac{\mathrm{e}^{-\mathrm{i}\Phi(\vec{k},t)}}{\sqrt{\pi/2}}
\sum_{l,m}\mathrm{i}^{l}j_{l}(kr)Y_{l}^{m}(\Omega_{r})Y_{l}^{m*}(\Omega_{k}).
\label{eq:planewaveIntoSphericalHarmonics}
\end{equation}
With this, the commutator matrix element is
\begin{align}
&\l\chi_{\vec{k}}(t)|\Big[-\frac{\Delta}{2}+\mathrm{i}\vec{A}\cdot\vec{\nabla},\Theta\Big]|\psi(t)\r=
\nonumber\\&
\qquad\frac{\mathrm{e}^{\mathrm{i}\Phi(\vec{k},t)}}{\sqrt{\pi/2}}R_c^2\sum_{l,m}(-\mathrm{i})^{l}Y_{l}^{m}(\Omega_{k})
\left(J_{lm}+\mathrm{i}A_z K_{lm}\right)
\end{align}
with the usual flux for the $lm$ partial wave
\beq
J_{lm}:=\frac{1}{2}j'_{l}(kR_{c})R_{lm}(R_{c},t)
-\frac{1}{2}j_{l}(kR_{c})R'_{lm}(R_{c},t)
\eeq
and the correction term for the dipole field
\beq
K_{lm}:=\sum_{s=\pm1}
\langle Y_{l}^{m}|\cos\theta|Y_{l+s}^{m}\rangle j_{l}(kR_{c})R_{l+s,m}(R_{c},t)
\eeq

\section{Rydberg state averaging\label{sec:appen:averaging}}

\begin{figure}[H]
\includegraphics[width=0.5\columnwidth]{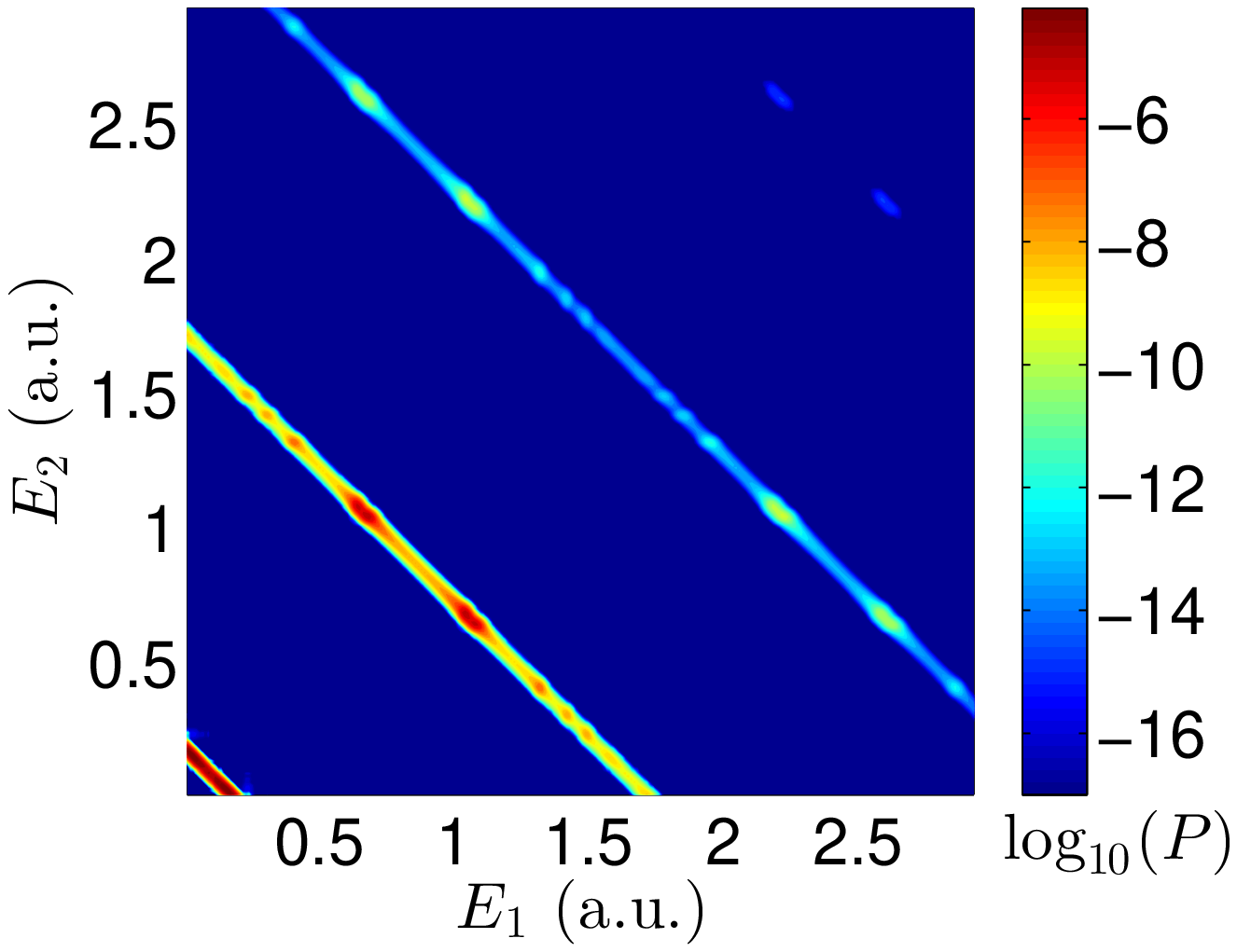}\includegraphics[width=0.5\columnwidth]{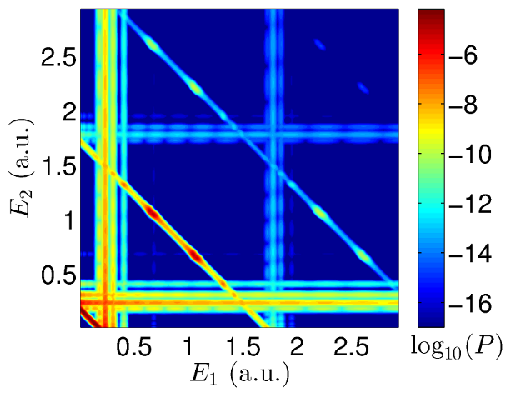}
\caption{\label{fig:noaveraging} %178345 228028
Left: Energy probability distribution $P(E_{1},E_{2})$ for a $n=120$
cycle pulse with photon energy $\hbar\omega=42\eV$ and peak intensity
$I=10^{13}\Wcm$ using a $\cos^{8}$-envelope and averaging time $\Delta T=1\ps$. 
Right: Without averaging.}
\end{figure}

Rydberg states extending beyond the tSurff radius $R_{c}$ lead to artificial
oscillating contributions in the photo-electron spectrum.
Assume that a hydrogen-like Rydberg state $\phi_n$ with energy $E_n<0$ and angular quantum number $l=n-1$
is populated at some time $T_0$ after the end of the pulse.
Its contribution to the tSurff spectrum 
\begin{equation}
b_n(k,T)\sim\langle \mathrm{e}^{\mathrm{i}\vk\vr}|\Theta |\phi_n \rangle \sim \frac{\sin(kR_c)}{k^2}.
\end{equation}
oscillates at constant amplitude
\begin{equation}
b_n(k,T')=b_n(k,T)\mathrm{e}^{\mathrm{i}(T'-T)(E_n-k^2/2)}.
\end{equation}
Averaging $b(k,T')$ over a time-interval $\Delta T$
\begin{equation}
\frac{1}{\Delta T}\int_{T}^{T+\Delta T} b_n(k,T')\mathrm{d}T'\sim \frac{b_n(k,T_0)}{\Delta T (E_n-k^2/2)}.
\end{equation}
suppresses the artifact.
In the spectra the decrease squares $\sim (\Delta T)^{-2}$.
True outgoing flux is not affected, if averging starts at times $T$ where all relevant flux has
passed the surface.
Fig.~\ref{fig:noaveraging} illustrates the suppression for the two particle case.

\section{Radial basis and inverse overlap\label{sub:Radial-Basis:-Continuity}}

The following was applied to the two radial axes
of the two electron system~(\ref{eq:discretization}) and the radial
axis of the single electron system~(\ref{eq:RaddiscOnS}). We use a
high degree finite element discretization whose construction is described
in~\cite{Scrinzi2010}. We divide the axis into $N$
elements $[r_{n-1},r_{n}]$, $n=1,\dots ,N$. On each element
we have $p_{n}$ linearly independent functions $\{f_{i}^{(n)},i=1\dots p_{n}\}$
satisfying
\begin{eqnarray}
f_{k}^{(n)}(r_{n-1})= & \,0\, & =f_{k}^{(n)}(r_{n})\nonumber\\
\text{except }f_{1}^{(n)}(r_{n-1})= & \,1\, & =f_{p_{n}}^{(n)}(r_{n}).
\end{eqnarray}
With these conditions, the overlap matrix on each finite element $S_{ij}^{(n)}=\langle f_{i}^{(n)}|f_{j}^{(n)}\rangle$
can be transformed such that its diagonal except for the two off-diagonal
elements $\langle f_{1}^{(n)}|f_{p_{n}}^{(n)}\rangle\neq0$. The end-elements $n=0$ and $n=N$
may have modified constraints for the purpose of implementing the boundary conditions.
The radial wavefunction is then given by $\psi(r,t)=\sum_{n=1}^{N}\sum_{k=1}^{p_{n}}f_{k}^{(n)}(r)c_{k}^{(n)}(t)$.
Continuity across element boundaries is assured by demanding $c_{p_{n-1}}^{(n-1)}=c_{1}^{(n)}$
for $n\geq2$. The application of an operator matrix thus amounts
to the blockwise application to each finite element and then enforcing
continuity, which is done by averaging the corresponding coefficients
\begin{equation}
\Big(c_{p_{n-1}}^{(n-1)},c_{1}^{(n)}\Big)\mapsto\Big(\frac{c_{p_{n-1}}^{(n-1)}+c_{1}^{(n)}}{2},\frac{c_{p_{n-1}}^{(n-1)}+c_{1}^{(n)}}{2}\Big).
\end{equation}
This map can be implemented by a projector $\mQ$ as follows.
First identify the wavefunctions with the coefficient vector $\psi\leftrightarrow\vec{c}\in\mathbb{C}^{d}$,
where $d=\sum_{n=1}^{N}p_{n}$ gives the overall number of coefficients.
For 
\begin{equation}
\vec{n}^T=(0,\dots,0,-\frac{1}{\sqrt{2}},0,\dots,0,\frac{1}{\sqrt{2}},0,\dots,0)\in\mathbb{R}^{d}\label{eq:discontinuityVector-1}
\end{equation}
the map $\mId-\vec{n}\otimes\vec{n}^{\mathrm{T}}$ realizes the above
averaging for the element boundary at $r_{n}$ if the two non-zero
coefficients in $\vec{n}$ are at the positions of $c_{p_{n-1}}^{(n-1)}$
and $c_{1}^{(n)}$ in the overall coefficient vector $\vec{c}$. The
full projector is then given by $\mQ=\widehat{\id}-\sum_{n}\vec{n}\otimes\vec{n}^{\mathrm{T}}$,
which can be written as 
\begin{equation}
\mQ=\mId-\mR\mR^{\mathrm{T}}
\end{equation}
with the $d\!\times\!(N-1)$ matrix
\begin{equation}
\mR=\Big(\vec{2},\dots,\vec{n},\dots,\vec{N}\Big).
\end{equation}
The application of the inverse overlap requires extra attention as it 
must be constrained to the subspace of continuous vectors $\mQ\vc$: 
it is the inverse in the sense $(\mQ\mS\mQ)^{-1}(\mQ\mS\mQ)=(\mQ\mS\mQ)(\mQ\mS\mQ)^{-1}=\mQ$.
The
formula used is a variation of the Woodbury matrix identity. It takes the form
\begin{equation}
(\mQ\mS\mQ)^{-1}  =  \left[\mId-\mS^{-1}\mR\big(\mR^{\mathrm{T}}\mS^{-1}\mR\big)^{-1}\mR^{\mathrm{T}}\right]\mS^{-1},
\label{eq:Sinverse_formula}
\end{equation}
where $\mS^{-1}$ is the blockwise inverse without the constraint $\mQ$.
This reduces the application of inverse of $\mQ\mS\mQ$ 
to the application of $N$ near diagonal blockwise inverse $[\mS\up{n}]^{-1}$ (operations count $\propto Np$)
and the correction by $\mS^{-1}\mR\big(\mR^{\mathrm{T}}\mS^{-1}\mR\big)^{-1}$
with operations count $\sim dN+N^2$. 

In two particle calculations two one-particle inverses need to be applied in sequence because of the tensor product form of the overlap
$\mS=\mS_1\otimes\mS_2$.

\section{Multipole operators\label{sec:app:multipole}}

In the multipole expansion of the electron-electron interaction (section~\ref{sec:e-e-int}) 
appear multipole matrices $\mV\up{\la}$.
For fixed angular momentum indices (omitted here)
and fixed finite element $(n_1,n_2)$,
and inserting the radial basis, Eq.~(\ref{eq:radialProductR}), these are given by
\bea
\lefteqn{
\mV\up{\la,n_1n_2}_{p_1'p_2',p_1p_2}=\int \!\mathrm{d}r_1 \int \!\mathrm{d}r_2\, \frac{\min(q_i,q_j)^\la}{\max(q_i,q_j)^{\la+1}}
}
\nonumber\\&& 
\qquad\times
f\up{n_1}_{p_1'}(r_1)f\up{n_2}_{p_2'}(r_2)
f\up{n_1}_{p_1}(r_1)f\up{n_2}_{p_2}(r_2).
\label{eq:Vlambda}
\eea
If the maximal degree of the polynomial expansion is $P-1$, then the product polynomials 
\[
f\up{n_i}_{p_i'}(r_i)f\up{n_i}_{p_i}(r_i)=:F\up{n_i}_{K_i}(r_i),\quad i=1,2
\] 
have maximal degree $2P-2$ with at most $2P-1$
linearly independent functions $F\up{n_i}_{K_i}(r_i)$. We rearrange the indices
$p_1'p_2',p_1p_2$ into $K_1=(p_1'p_1)$ and $K_2=(p_2'p_2)$ 
and consider $\mV\up{\la}_{p_1'p_2',p_1p_2}$ as a two-index $P^2\!\times\! P^2$ matrix~$\mW\up{\la}_{K_1,K_2}$. 
This matrix has a maximal rank of $R:=2P-1$ and, in a suitable representation,
it reduces to a $R\!\times\!R$ 
%(or a diagonal $R^2\!\times\!R^2$) 
matrix $\mD\up{\la}_{ij}$. One such representation is with 
respect to $R$-point Gaussian quadrature grids:
$\{q\up{n_1}_i,w\up{n_1}_i\}_{i=1\dots R}$ on the the interval $[r_1\up{n_1-1},r_1\up{n_1}]$
and $\{q\up{n_2}_j,w\up{n_2}_j\}_{j=1\dots R}$ on $[r_2\up{n_2-1},r_2\up{n_2}]$, where
$w\up{n_1}_i$ and $w\up{n_2}_j$ are the associated quadrature weights.
For these grids we have
\beq
\mD\up{\la}_{ij}=\sum_{K_1,K_2}  F\up{n_1}_{K_1}(q_i\up{n_1})\mW\up{\la}_{K_1,K_2}F\up{n_2}_{K_2}(q_j\up{n_2})
\label{eq:Dlambda}.
\eeq
%As all $F\up{n_i}_{K_i}(r_i)$ are polynomials of maximal degree $2P-2$, they are fully defined by their values
%on the $R=2P-1$ distinct quadrature points.
The transformation from the $P^2$ coefficients to the $R<P^2$ coefficients
can be done separately for each coordinate,
i.e.\ it has tensor product structure and
operations count $RP(R+P)\propto P^3$.
It is given by $\mT\up{n_1}\otimes\mT\up{n_2}$ with
\begin{equation}
\mT_{i_\al,p_\al}\up{n_\al} = \sqrt{w\up{n_\al}_{i_\al}}f\up{n_\al}_{p_\al}(q\up{n_\al}_{i_\al}),\quad \al=1,2.
\end{equation}
%This transformation and also the back-transformation are exact, as the 
%$R$-point Gaussian quadrature is exact up to a maximal polynomial degree $4P-3$, which exceeds
%all polynomial degrees appearing here. 
Thus, the application of $\mV\up{\la}_{p_1'p_2',p_1p_2}$ amounts to a transformation
to the reduced representation with $R$ coefficients,
the coefficient-wise multiplication with $\mD\up{\la}_{ij}$ (operations count $R^2$), 
and the back-transformation to the representation with $P^2$ coefficients:
\bea
\lefteqn{\mV\up{\la,n_1n_2}_{p_1'p_2',p_1p_2} =}\nonumber
\\&&\quad \sum_{ij}
\left(\mT_{i,p'_1}\up{n_1}\otimes\mT_{j,p'_2}\up{n_2}\right)^\mathrm{T}
\mD\up{\la}_{ij}
\left(\mT_{i,p_1}\up{n_1}\otimes\mT_{j,p_2}\up{n_2}\right),
\eea
compare Eq.~(\ref{eq:e-e-int-diagonal}).

The integrals $\mW\up{\la}_{K_1,K_2}$ for
generating the correct $\mD_{ij}\up{\la}$~(\ref{eq:Dlambda}) need to be evaluated only once
during setup. This step must not be bypassed by using 
$\min(q_i,q_j)^\la/ \max(q_i,q_j)^{\la+1}$, as this potential is not suitable 
for direct integration with a Gaussian quadrature on the product grid $q_iq_j$. 

In practice, we found that the quadratures do not need to be done exactly.
Minor quadrature errors introduced by a Gaussian quadrature grid with only $P$ or even fewer points
are acceptable,
which further reduces the operations count.

It is obvious from the derivation, that the same procedure can be applied for any 
two-dimensional multiplicative potential and gives the exact matrix elements for a given
polynomial product basis. It is most useful for potentials that have points of non-analyticity,
such as the Coulomb potential. For potentials with a convergent Taylor series, Gaussian
quadrature can be usually applied directly.

\section{irECS\label{sec:appen:absorption}}
ECS is the coordinate rotation into the lower
complex plane, defined by
\begin{equation}
r\mapsto r_{\theta}=\begin{cases}
r & r\leq R_{0}\\
\mathrm{e}^{\mathrm{i}\theta}(r-R_{0})+R_{0} & r>R_{0}.
\end{cases}
\end{equation}
with the scaling angle $\theta>0$. We can choose any $R_0>R_c$.
We let the scaling radius $R_{0}$ fall onto an element boundary of the
finite element discretization of the radial axis.
Following the specifications in~\cite{Scrinzi2010}, ECS can then be realized
by introducing an explicit discontinuity in the basis functions $f_{i}$
at the scaling radius $R_{0}$: 
\begin{equation}
f_{i}^{(\theta)}(r)=\begin{cases}
f_{i}(r) & r < R_{0}\\
\mathrm{e}^{\mathrm{i}\theta/2}f_{i}(r) & r>R_{0}.
\end{cases}
\end{equation}
This translates into scaled matrices in the discretized TDSE (\ref{eq:discretizedTDSE}).

The overlap matrix $\mS_{ij}=\langle f_{i}|f_{j}\rangle$ transforms
into $\mS_{\theta}$ in the scaled region as
\begin{eqnarray}
\mS_{\theta,ij} & = & \int\mathrm{d}r\,\big(\mathrm{e}^{\mathrm{i}\theta/2}f_{i}\big)(r)\big(\mathrm{e}^{\mathrm{i}\theta/2}f_{j}\big)(r)=\mathrm{e}^{\mathrm{i}\theta}\mS_{ij}.\label{eq:ScomplesScaled}
\end{eqnarray}
Note that the left hand side $\mathrm{e}^{\mathrm{i}\theta/2}$ does not get
complex conjugated, as explained in~\cite{Scrinzi2010}. The various
terms in the Hamiltonian matrix $\mH$ transform accordingly.
Potential terms are evaluated at complex values, which requires the
analytic continuation of these potentials
\begin{equation}
\mV_{\theta,ij}=\langle f_{i}\up{\th}\big|V(r_{\theta})|f_{j}\up{\th}\rangle.\label{eq:potentialComplexScaled}
\end{equation}
The same principle applies for the evaluation of derivatives, observing that 
$\partial/\partial(\mathrm{e}^{\mathrm{i}\th}r)
=\mathrm{e}^{-\mathrm{i}\th}\partial/\partial r$.
For example, for elements outside $R_0$, the complex scaled matrix for the second derivative is
\begin{equation}
-\left\l \frac{\partial f\up{\th}_i}{\partial \mathrm{e}^{\mathrm{i}\th}r}|\frac{\partial f\up{\th}_i}{\partial \mathrm{e}^{\mathrm{i}\th}r}\right\r=
-\mathrm{e}^{-\mathrm{i}\th}\left\l \frac{\partial f_i}{\partial r}|\frac{\partial f_j}{\partial r}\right\r.
\end{equation}
The implementation of complex scaling thus amounts to simple multiplications
of the matrices with factors of $\mathrm{e}^{\mathrm{i}\theta}$, and
evaluation of potential terms at complex values.\\
This applies to both standard ECS and infinite range ECS. 
For the implementation of irECS basis functions on the last element
extend to infinity and integration needs to be performed over the 
complete range~\cite{Scrinzi2010}. %

\bibliographystyle{apsrev}% APS bibliography style. use only with .bib that has no URL specified.
\bibliographystyle{unsrt}
\bibliographystyle{IEEEtran} % for links in references
\bibliography{DIpaperBib}

\end{document}

%% file: my_commands.tex
\usepackage{amsmath}
\usepackage{amssymb}
\usepackage{makeidx}
\usepackage{graphicx}
\usepackage{color}

\newcommand{\wrt}{w.r.t.\ }

\newcommand{\hide}[1]{}

\newcommand{\up}[1]{ ^{(#1)}}
\newcommand{\inv}[1]{ ^{-#1}}

\newcommand{\nm}{{\rm\,nm}}

\newcommand{\fs}{{\rm\,fs}}
\newcommand{\Wcm }{ \, \mathrm{W} \! / \mathrm{cm}^2 }
\newcommand{\au}{ \, \mathrm{a.u.}}
\newcommand{\ps}{ \, \mathrm{ps}}
\newcommand{\eV}{ \, \mathrm{eV}}

\newcommand{\myvec}[1]{\vec{#1}}
\renewcommand{\vr}{{\myvec{r}}}

\newcommand{\va}{{\myvec{a}}}
\newcommand{\vc}{{\myvec{c}}}
\newcommand{\vd}{{\myvec{d}}}

\newcommand{\vk}{{\myvec{k}}}

\newcommand{\vb}{{\myvec{b}}}
\newcommand{\vA}{{\myvec{A}}}

\newcommand{\vC}{{\myvec{C}}}

\newcommand{\id}{{1\hspace{-0.85ex}1}}

\newcommand{\mymat}[1]{{\widehat{#1}}}

\newcommand{\mH}{\mymat{H}}

\newcommand{\mD}{\mymat{D}}

\newcommand{\mP}{\mymat{P}}
\newcommand{\mQ}{\mymat{Q}}
\newcommand{\mR}{\mymat{R}}

\newcommand{\mS}{\mymat{S}}
\newcommand{\mT}{\mymat{T}}

\newcommand{\mV}{\mymat{V}}
\newcommand{\mW}{\mymat{W}}

\newcommand{\mId}{\mymat{\id}}

\newcommand{\cA}{\mathcal{A}} % anti-symmetrization
\renewcommand{\r}{\rangle}
\renewcommand{\l}{\langle}

\newcommand{\ddt}{\frac{\mathrm{d}}{\mathrm{d}t}}

\newcommand{\om}{\omega}

\newcommand{\al}{\alpha}
\newcommand{\la}{\lambda}

\newcommand{\be}{\beta}

\renewcommand{\th}{\theta}

\newcommand{\lcase}{\left\{\begin{array}{ll}}
\newcommand{\rcase}{\end{array}\right.}
\renewcommand{\bar}{\begin{array}{ll}}
\newcommand{\ear}{\end{array}}
\newcommand{\bma}{\begin{pmatrix}}
\newcommand{\ema}{\end{pmatrix}}
\newcommand{\beq}{\begin{equation}}
\newcommand{\eeq}{\end{equation}}
\newcommand{\bel}[1]{\begin{equation}\label{eq:#1}}
\newcommand{\eel}{\end{equation}}
\newcommand{\bea}{\begin{eqnarray}}
\newcommand{\eea}{\end{eqnarray}}
\newcommand{\beaNN}{\begin{eqnarray*}}
\newcommand{\eeaNN}{\end{eqnarray*}}

\newcounter{lecture}

\newcommand{\vna}{\vec{\nabla}}